\documentclass[11pt,amsmath,floats,floatfix,nofootinbib,superscriptaddress]{revtex4}
\usepackage{graphicx} \usepackage{bm} \usepackage{epsfig}
\usepackage{pstricks}
\usepackage{setspace}
\usepackage{graphicx}
\usepackage{bm}
\usepackage{epsfig}
\usepackage{color}
\usepackage{colordvi}
\newcommand{\beq}{\begin{equation}}
\newcommand{\eeq}{\end{equation}}
\newcommand{\bea}{\begin{eqnarray}}
\newcommand{\eea}{\end{eqnarray}}
\newcommand{\bit}{\begin{itemize}}
\newcommand{\eit}{\end{itemize}}

\newcommand{\nn}{\nonumber}

\begin{document}

\title{Spin induced multipole moments for the gravitational wave flux from binary inspirals to third Post-Newtonian order}
\author{Rafael A. Porto}
\email{rporto@ias.edu}
\affiliation{Kavli Institute for Theoretical Physics, University of California, Santa Barbara CA 93106, USA}
\affiliation{School of Natural Sciences, Institute for Advanced Study, Olden Lane, Princeton, NJ 08540, USA}
\affiliation{Department of Physics and ISCAP, Columbia University, New York, NY 10027, USA}
\author{Andreas Ross}
\email{andreasr@andrew.cmu.edu}
\affiliation{Yale University Dept. of Physics, New Haven CT 06511, USA}
\affiliation{Carnegie Mellon University Dept. of Physics, Pittsburgh PA 15213, USA}
\author{Ira Z. Rothstein}
\email{izr@andrew.cmu.edu}
\affiliation{Carnegie Mellon University Dept. of Physics, Pittsburgh PA 15213, USA}

\begin{abstract}
Using effective field theory techniques we calculate the source multipole moments needed to obtain the spin contributions to the power radiated in gravitational waves from inspiralling compact binaries to third Post-Newtonian order (3PN). The multipoles depend linearly and quadratically on the spins and 
include both spin(1)spin(2) and spin(1)spin(1) components. The results in this paper provide the last missing ingredient required to determine the phase evolution to 3PN including all spin effects which we will report in a separate paper.
\end{abstract}
\date{\today}

\maketitle

\newpage

\tableofcontents

\section{Introduction}

Coalescing compact binary systems are one of the most promising classes of sources in the search for gravitational waves through direct detection at earth based interferometers such as LIGO \cite{Abramovici:1992ah}, VIRGO \cite{Giazotto:1988gw} and at the planned LISA \cite{Danzmann:2003tv} and Einstein Telescope \cite{etele} experiments.  One of the main goals of these instruments, beyond wave detection itself, is to extract masses and spins from the signals. However, the ability to perform these
extractions is predicated upon precise theoretical predictions.

During the initial inspiral phase of the binary coalescence the compact objects are well separated and one can study their dynamics analytically in the post-Newtonian (PN) approximation. For binary systems the PN approximation is a non-relativistic expansion in the typical three-velocities $v$ in the binary, and the standard lore is that accurate parameter extraction requires the knowledge of the phase of the gravitational wave signal at least to 3PN\footnote{We follow the usual convention in the literature and label leading effects as 0PN with higher order corrections counted relative to these. For example, the leading conservative dynamics gives the Newtonian or 0PN equations of motion, the leading energy flux in gravitational waves due to the leading spinless mass quadrupole (0PN) of the system is called the 0PN flux, and the phase evolution which follows from the 0PN conservative dynamics and flux is referred to as the 0PN phase.} or $\mathcal O(v^6)$ \cite{Cutler:1992tc, cutler, poisson}. For spinless systems 3.5PN precision has been reached in the last decade by a series of dedicated calculations, see \cite{Blanchet:2002av} for a review and further references.\\

Recent measurements of black hole spins indicate that black holes in binary systems may frequently be close to maximally rotating \cite{McClintock:2009dn}, and that spin effects have a significant impact on the development of accurate gravitational wave templates \cite{pan, buo3, hughes,arun}. In addition, higher order PN corrections including spin are relevant for comparison between analytic results and numerical simulations \cite{lousto,lousto2}. It is therefore desirable  to account for spin effects at the same degree of accuracy as the spin-independent components.

Currently, the phase evolution of the expected gravitational wave signal is known to 2.5PN precision when including spin (in a power counting assuming maximally rotating black holes\footnote{The spin $\mathbf S$ of a compact object is related to its mass $m$ by $|\mathbf S | = a G m^2$ where $G$ is Newton's constant and $0 \leq a \leq 1$. Assuming maximal rotation means we power count $a \sim 1$, but our results hold of course for arbitrary $a$. For binary systems in the PN regime and for our purposes the spin scaling is conveniently written as $|\mathbf S | \sim m r v^2 \sim L v$ where $r$ is the orbital separation between the binary constituents and $L$ the orbital angular momentum of the binary.}). This includes the leading order (LO) spin-orbit terms, i.e. effects linear in spin, at 1.5PN \cite{kidder1,kidder2}, the LO spin(1)spin(2) and spin(1)spin(1) contributions at 2PN \cite{kidder1, kidder2, Mikoczi:2005dn} and the next-to-leading order (NLO) spin-orbit effects at 2.5PN \cite{owen,buo1,buo2}. \\

In this paper we calculate all missing ingredients to obtain the spin contributions to the energy flux and the phase evolution to 3PN order using the effective field theory (EFT) techniques first introduced in \cite{nrgr,nrgr1}, and extended in \cite{nrgrs,nrgrproc} to include spin degrees of freedom. We include the following multipole moments: The mass quadrupole with ${\cal O}({\bf S}_A)$ and ${\cal O}({\bf S}^2_A)$ components to NLO as well as the leading ${\cal O}({\bf S}_A {\bf S}_B)$ terms.  The current quadrupole with LO ${\cal O}({\bf S}^2_A)$ and up to NLO ${\cal O}({\bf S}_A)$ contributions. The ${\cal O}({\bf S}_A)$ and ${\cal O}({\bf S}_A^2)$ terms for the mass octupole at LO.  The LO effects of ${\cal O}({\bf S}_A)$ in the current octupole, and  the spin-independent 1PN corrections to the current quadrupole moment. 

While the components of ${\cal O}({\bf S}_A)$ have been computed at the same order but with different conventions in \cite{buo2}, the NLO ${\cal O}({\bf S}^2_A)$ as well as the ${\cal O}({\bf S}_A {\bf S}_B)$ terms are new results. Together with the conservative part of the dynamics obtained in \cite{eih,comment,nrgrss,nrgrs2,nrgrso} using the same conventions as in this work, our results enable us to compute all spin-dependent components of the energy flux and thus the phase up to 3PN. These will be reported in a separate publication \cite{powerloss} where we will also include a comparison with the 2.5PN results of \cite{buo2}.\\

Our computation is performed within the framework of Non-Relativistic General Relativity (NRGR) \cite{nrgr}. The EFT treatment (see \cite{iraeft} for a review of EFT methods) of binary inspirals is based on the separation of the physical scales in the binary which are disentangled and treated one at a time. This is achieved by introducing  a nontrivial decomposition of the gravitational field into different kinematic regions and by setting up several EFTs in stages removing each scale. This makes NRGR a systematic and simple tool to perform calculations in the PN expansion, see \cite{nrgrLH} for a pedagogical review.

The original paper proposing NRGR \cite{nrgr} contained the computation of the spinless 1PN conservative dynamics and the LO radiated power which is due to LO spinless mass quadrupole radiation. Dissipative effects were included in \cite{dis1} via the introduction of additional worldline degrees of freedom. The EFT framework works very efficiently in higher order PN calculations and it was used in \cite{andi1} to compute the spinless conservative part of the binary dynamics to 2PN or NNLO using the convenient parameterization of the gravitational field of \cite{Kol:2007bc}.

The EFT for the radiation sector has been developed further in \cite{rad1} where radiative corrections to gravitational wave emission such as tail effects were considered including a proper treatment of all divergences. The power emitted in gravitational waves from spinless binaries was reproduced at 1PN, 1.5PN and 2.5PN, and the leading ultraviolet logarithms in the gravitational wave power which arise at 3PN were computed and resummed using the renormalization group \cite{rad1}. In order to calculate real-time quantities such as the gravitational waveform or the self-force in a self-consistent framework, NRGR was translated to the in-in formalism in \cite{chad2}.

NRGR was extended to include spin degrees of freedom in \cite{nrgrs}, where the LO spin-spin and spin-orbit effects were reproduced as a sum of very simple Feynman diagrams. The NLO spin(1)spin(2) potential was then first derived using the Newton-Wigner (NW) spin supplementarity condition (SSC) at the level of the action in \cite{eih,nrgrproc,comment}. Afterwards in \cite{nrgrss}  these results were re-derived using a Routhian approach in the covariant SSC, which was also used to compute the NLO spin(1)spin(1) potential including finite size effects \cite{nrgrs2}. These results were independently corroborated in \cite{Schafer3pn,Schafer3pn2} using a more traditional  set of techniques, and complete agreement on the conservative 3PN spin-spin dynamics has been reported in \cite{Hergt:2010pa}. The NLO spin-orbit potential was recently derived using the EFT approach in \cite{nrgrso, Perrodin:2010dy,Levi:2010zu} and shown to agree  \cite{nrgrso} with the results in \cite{buo1}. Dissipative effects due to spin were also computed in \cite{dis2}.
 
The worldline EFT techniques of \cite{nrgr} are clearly not limited to the PN regime of binary inspirals, or only to gravitational interactions, and can be applied to other systems with separate scales where a sensible power counting can be established. Inspired by \cite{nrgr}, these techniques have been used to study the electromagnetic self-force on extended objects \cite{adam}, the radiation-reaction force in the extreme mass ratio limit \cite{chad1} and the thermodynamics of caged black holes \cite{cbh,Kol:2007rx,andi2}. More recently, an EFT setup has been proposed to study cosmological perturbations \cite{eftfluid}.\\

This paper is organized as follows. In Sec. \ref{sec:setup} we summarize the EFT setup for radiation effects from spinning extended objects. The matching calculation for the source multipole moments is performed in Sec. \ref{sec:match}. Our combined results for the multipole moments are presented in Sec. \ref{sec:total} where we outline the assembly of all ingredients to the 3PN spin effects in the radiated power. The main results of this paper are collected in Eqs. (\ref{Qijtot} - \ref{octsless}) written in the covariant SSC. The energy flux and phase evolution will be reported in a separate publication \cite{powerloss}. We conclude in Sec \ref{sec:concl}. The Appendices contain some subtleties regarding spin effects in GR and the spin-independent multipole moments computed with our set of conventions needed for the power loss to 3PN.

\section{Effective Field Theory Setup}\label{sec:setup}

The physics of a compact binary system is governed by three relevant scales, the size of the compact objects $R$, their separation $r$ and the wavelength $\lambda$ of the gravitational waves it emits. In the PN regime all three are well separated and their hierarchy $R \ll r \ll \lambda$ is controlled by powers of the small expansion parameter $v$.  We have the relations $R \sim r v^2$ and $\lambda \sim r/v$ where the former assumes compact objects (such as black holes).

The NRGR construction of EFTs describing binaries proceeds in stages. Above the scale $R$ we describe the compact binary system as two point particles coupled to gravity with higher dimensional operators\footnote{The term ``operator" is used here only to be consistent with terminology developed for general EFTs.} encoding finite size effects. This EFT is valid 
in a general frame, and can be written down in a generally covariant fashion. The corresponding action can be utilized in either the extreme mass ratio or PN scenarios. 

In order to properly disentangle the scales we decompose the gravitational field into different momentum regions which are responsible for different physical effects. Hence $h_{\mu \nu}/m_p \equiv g_{\mu \nu} - \eta_{\mu \nu}$ (where the Planck mass $m_p$ is related to Newton's constant by $G=1/(32\pi m^2_p)$ in our conventions) is written as the sum of potential modes $H_{\mu \nu}$ and radiation modes $\bar h_{\mu \nu}$ with momentum scalings $\partial_\alpha H_{\mu \nu} \sim (v/r, 1/r) H_{\mu \nu}$ and $\partial_\alpha \bar h_{\mu \nu} \sim (v/r, v/r) \bar h_{\mu \nu}$ respectively \cite{nrgr}. The potential modes are responsible for the binding dynamics of the binary whereas the on-shell radiation modes describe the gravitational waves which propagate out to the detector. Moreover, the radiation field $\bar h_{\mu \nu}$ has to be Taylor expanded around a single point in order to achieve a uniform power counting at the level of the action \cite{multip}. By working in the background field gauge one can preserve diffeomorphism invariance in the radiation sector of the theory, and furthermore the potential and radiation modes can be gauge fixed independently\footnote{There are no double counting issues between modes in the EFT due to the differing analytic structure of their respective propagators in momentum space (see \cite{iraeft} for details.)}.\\

To compute radiation observables, such as the energy flux, we ``integrate out'' (i.e. solve for and plug their solution back into the action) the potential modes. The resulting radiation theory describes the binary as a single point particle coupled to gravity and endowed with a set multipole moments which describe its internal dynamics. 

In the next  section we review the basic aspects of the EFT approach. Readers interested in further details should consult \cite{nrgr,nrgrs,rad1}.

\subsection{The Radiation Sector of NRGR}

 The EFT describing the radiation couplings of the binary system is naturally formulated as a multipole expansion, since it corresponds to a long wavelength approximation. It is convenient to work in the center-of-mass frame where the action of the radiation EFT \cite{rad1} reads
\begin{align}
S_{\rm rad}[\bar h] = & - \int dt \left(m\sqrt{g_{00}} + \frac{1}{2} L^{ij} \omega_{ij0} \right) \notag \\
& + \frac{1}{2} \int dt \left(I^{ij} E_{ij} - \frac{4}{3} J^{ij} B_{ij} + \frac{1}{3} I^{ijk} \nabla_k E_{ij} - \frac{1}{2} J^{ijk} \nabla_k B_{ij} + \cdots \right). \label{eq:multipoleAction}
\end{align}
The first two terms contain the couplings to the total mass $m$ and the total angular momentum $L^{ij}$ which do not radiate since they are conserved. However, these terms source the Kerr background due to the binary itself in which the emitted gravitational waves propagate. Note also that our definitions of mass $m$ and angular momentum $L^{ij}$ are equivalent to the ADM prescription.
The source multipole moments which can radiate are of either electric or magnetic type parity, and couple to the electric ($E_{ij}$) or magnetic ($B_{ij}$) components of the Weyl tensor and derivatives thereof.

\begin{figure}[t]
   \centering
   \includegraphics[width=5cm]{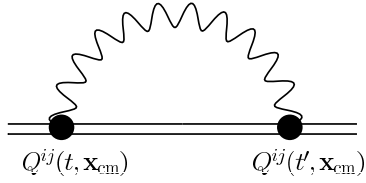}
\caption{\sl Leading order contribution to the effective action after integrating out the radiation field. The double solid line denotes the binary system and the wiggly line denotes the radiation graviton.}\label{qq}
\end{figure}

The observable we are ultimately interested in, the power loss or energy flux in gravitational waves, can be either extracted from the imaginary part of  `vacuum energy' diagrams such as the one shown in Fig. \ref{qq} or from squaring graviton emission amplitudes and summing over polarizations. Either way yields the standard result
\begin{equation}
{\dot E}  =   \frac{G}{5} \left< \left({d^3\over  dt^3} I^{ij}(t)\right)^2 \right>+ \frac{16 G}{45} \left< \left({d^3\over  dt^3} J^{ij}(t)\right)^2 \right>+  \frac{G}{189} \left< \left({d^4\over  dt^4} I^{ijk}(t)\right)^2 \right> +  \frac{G}{84} \left< \left({d^4\over  dt^4} J^{ijk}(t)\right)^2 \right> + \cdots\label{powerloss} 
\end{equation}
where we did not include effects in the radiation theory due to the nonlinearities of gravity such as the tail effect.

The information about the binary and its internal dynamics is encoded in the Wilson coefficients of the action in Eq. (\ref{eq:multipoleAction}), i.e. in the multipole moments. These are obtained through a matching calculation which is the central part of this work.  For this purpose we compare the action for the radiation EFT in Eq. (\ref{eq:multipoleAction}) with the EFT valid below the orbital scale $r$, but above the finite size scale $R$, which describes the binary as two point particles coupled to both potential and radiation modes of the gravitational field. We then calculate the one-graviton emission amplitude from Feynman diagrams with a single external radiation field and arbitrary potential exchanges between the two point particles. This amplitude can be expressed in the form of an effective action linear in the radiation field $\bar h_{\mu \nu}$
\beq
\label{Srad}
\Gamma[\bar h] = -\frac{1}{2m_p} \int d^4x T^{\mu\nu}(x){\bar h}_{\mu\nu}(x), 
\eeq
where $\partial_\mu T^{\mu\nu}=0$ due to the gravitational Ward identity. Eq. (\ref{Srad}) serves as our definition of the stress energy pseudotensor $T^{\mu\nu}$ which contains gravitational contributions due to potential modes. 

In order for this effective action to obey a manifest power counting the radiation field ${\bar h}_{\mu\nu}(x)$ must be Taylor expanded around a single point within the binary \cite{multip} which is conveniently chosen to be the center-of-mass ${\bf x}_{\rm cm}=0$. Plugging this Taylor expansion into Eq. (\ref{Srad}) and organizing the multipole moments into irreducible representations of the rotation group using symmetric trace-free (STF) tensors (plus relations which follow from $\partial_\mu T^{\mu\nu}=0$ as well as the on-shell equations of motion), we can bring Eq. (\ref{Srad}) into the same multipole expansion form as the action for the radiation EFT of Eq. (\ref{eq:multipoleAction}). This procedure yields the matching for the multipole moments needed in terms of moments of $T^{\mu\nu}$. At the order we are working, the following expressions for the multipole moments are sufficient
\bea
I^{ij}_{0}&=&\int d^3{\bf x} \, T^{00} [{\bf x}^i{\bf x}^j]_{\rm TF}, \label{Q0ij}\\
I^{ij}_{1}&=&\int d^3{\bf x} \left(T^{ll}-\frac{4}{3}\dot T^{0l}{\bf x}^l+\frac{11}{42} \ddot T^{00}{\bf x}^2\right) [{\bf x}^i{\bf x}^j]_{\rm TF}, \label{Q1ij}\\
I^{ij}_2&=&\int d^3{\bf x} \left(\frac{2}{21} \ddot T^{ll} {\bf x}^2 + \frac{1}{6} \ddot T^{lm} {\bf x}^l {\bf x}^m - \frac{1}{7} \dddot T^{0l} {\bf x}^l {\bf x}^2 + \frac{23}{1512} \ddddot T^{00} {\bf x}^4 \right) \left[{\bf x}^i {\bf x}^j\right]_{\rm TF}, \label{Q2ij}\\
I_0^{ijk}&=& \int d^3{\bf x} \, T^{00} [{\bf x}^i{\bf x}^j{\bf x}^k]_{\rm TF},\label{Q0ijk} \\
I_1^{ijk}&=& \int d^3{\bf x} \left(T^{ll} - \dot T^{0l} {\bf x}^l + \frac{1}{6} \ddot T^{00} {\bf x}^2 \right) \left[{\bf x}^i {\bf x}^j {\bf x}^k\right]_{\rm TF}\label{Q1ijk} \\
J^{ij}_0 &=&  \int d^3{\bf x} \left[\epsilon^{ilk}T^{0k}{\bf x}^j{\bf x}^l \right]_{\rm STF} \label{J0ij}\\
J^{ij}_1&=& \int d^3{\bf x} \left[ \epsilon^{ilk} \left(\frac{1}{28} \dot T^{lm}{\bf x}^j{\bf x}^k{\bf x}^m + \frac{3}{28}  \dot T^{jk}{\bf x}^l{\bf x}^2\right) \right]_{\rm STF}, \label{J1ij}\\
J^{ijk}_0&=&  \int d^3{\bf x} \left[\epsilon^{iml} T^{0l} {\bf x}^m{\bf x}^j{\bf x}^k\right]_{\rm STF},   \label{Jijk}
\eea
where the STF prescription only acts on the free indices of the multipole moments.
Note that the couplings of the multipole moments which can radiate, quadrupole and higher, are all contained in the $\bar h_{ij}$ part of the effective action in Eq. (\ref{Srad}) since these describe the physical on-shell modes, i.e. the gravitational waves. Thus, in obtaining Eqs. (\ref{Q0ij}-\ref{Jijk}) we may Taylor expand only the radiation field $\bar h_{ij}$ and set $\bar h_{00}$ and $\bar h_{0i}$ to zero.

In computing the stress energy (pseudo)tensor which enters in Eqs. (\ref{Q0ij} - \ref{Jijk}) it is convenient to consider the partial Fourier transform
\beq
\label{partT}
T^{\mu\nu}(t,{\bf k}) = \int d^3{\bf x} T^{\mu\nu}(t,{\bf x}) e^{-i {\bf k}\cdot {\bf x}}
\eeq
since the calculation of Feynman diagrams is most easily performed in momentum space.
In particular, we can read off $T^{\mu\nu}(t,{\bf k})$ by calculating all diagrams
with one external radiation graviton leg.  By formally Taylor expanding 
\beq
\label{fourier}
T^{\mu\nu}(t,{\bf k}) = \sum_{n=0}^{\infty} \frac{(-i)^n}{n!} \left(\int d^3{\bf x} T^{\mu\nu}(t,{\bf x}) {\bf x}^{i_1}\cdots {\bf x}^{i_n}\right) {\bf k}_{i_1}\cdots {\bf k}_{i_n}
\eeq 
we can extract the moments needed  in Eqs. (\ref{Q0ij} - \ref{Jijk}) by direct comparison with the results from the diagram. For further details on the radiation EFT and the matching see \cite{rad1}.

\subsection{Spin Couplings}

The main advantage of our method is the introduction of an action from the onset. A spinless point particle of mass $m$ is described by the familiar worldline action
\begin{equation}
 S_{\rm pp} = - m \int d \tau \label{eq:WLnoSpin}
\end{equation}
with $d\tau = \sqrt{g_{\mu \nu} dx^\mu dx^\nu}$, and spin is added as an additional worldline degree of freedom. Spin couplings were introduced in the form of an action in \cite{nrgrs,eih,comment} or a Routhian in \cite{nrgrproc,nrgrss,nrgrs2}. In both cases the leading spin-graviton coupling is written as
\begin{equation} \label{act2}
S_{\rm sg} =  -\frac{1}{2} \int S^{ab}\omega_{ab\mu} u^\mu dt,
\end{equation}
in terms of the spin in the locally flat frame $S^{ab} \equiv S^{\mu\nu}e_{\mu}^a e_{\nu}^b$ and the spin connection $\omega_{ab\mu} \equiv e_{\nu b} D_\mu e^\nu_a$, where $e_{\mu}^b$ is a vierbein such that $e^b_\mu e^a_\nu \eta_{ab} = g_{\mu\nu}$. We chose $\lambda=x^0\equiv t$ parameterizing the worldline in terms of coordinate time, and therefore $u^\alpha\equiv \frac{d x^\alpha}{d \lambda} = v^\alpha\equiv \frac{d x^\alpha}{d t} = (1,{\bf v})$.

\begin{figure}[h!]
    \centering
    \includegraphics[width=9cm]{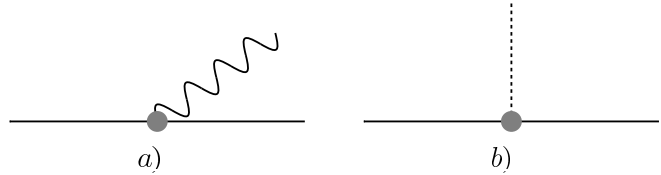}
\caption{\sl ${\cal O}({\bf S}_A)$ worldline coupling to: a) radiation and b) potential modes. A blob represents couplings linear in spin induced by the terms in $L_{\rm sg}$.}\label{world}
\end{figure}

Now we expand Eq. (\ref{act2}) in the weak gravity limit. We note that after expanding in powers of $h_{\mu\nu}$ we do not distinguish between local and covariant indices any more, but we consistently work with spin in the locally flat frame. We obtain at linear order in $h_{\mu\nu}$
\bea
L_{\rm sg}^0 &=& \frac{1}{2m_p}h_{i0,k}S^{ik},\label{sgnr1}\\
L^v_{\rm sg} &=& \frac{1}{2m_p}\left(h_{ij,k}S^{ik}v^j + h_{00,k}S^{0k}\right),\label{sgnr15}\\
L^{v^2}_{\rm sg} &=& \frac{1}{2m_p}\left(h_{0j,k}S^{0k}v^j +
h_{i0,0}S^{i0}\right)\label{sgnr2}, \\
L^{v^3}_{\rm sg} &=& \frac{1}{2m_p} h_{ij,0}S^{i0}v^j \label{sgnr25}, 
\eea
where we kept the counting of $v$ for potential modes of \cite{nrgrs}.
Time derivatives or factors of $S^{i0}(\sim S^{ij} v^j)$ both scale as relative factors of $v$. 
At ${\cal O}(h^2)$ we have the terms
\beq
L^{Sh^2}_{\rm sg} = \frac{1}{4m^2_p}S^{\alpha\beta} h^{\lambda}_\beta \left(\frac{1}{2} h_{\alpha\lambda,\mu} + h_{\mu\lambda,\alpha} -
h_{\mu\alpha,\lambda}\right)u^\mu .  \label{sgh2}
\eeq
Splitting $h_{\mu\nu}$ into the sum of radiation ($\bar h_{\mu\nu}$) and potential ($H_{\mu\nu}$) modes we obtain the Feynman rules linear in spin. The terms linear in the metric are depicted in Figs. \ref{world}$a$ for radiation and in \ref{world}$b$ for potential modes respectively.

\subsubsection{Spin Supplementarity Condition (SSC)}

The choice of SSC plays an important role in our analysis. Since we work with spin in the local frame, the SSC inevitably involves the particle's velocity in the local frame, and this introduces important effects in the dynamics \cite{comment, nrgrss} since we wish to express the velocity in the PN frame. For instance, to the order we are working in this paper the covariant SSC $S^{ab}p_b=0$ can be written as $S^{ab}e_{b\mu} u^\mu=0$, and we can express \cite{nrgrso}
\beq
\label{ssc1}
S_A^{i0}= \frac{1}{1+2(1-\kappa)\sqrt{1-{\bf v}_A^2}}\frac{\left(e^j_k({\bf x}_A) {\bf v}_A^k + e^j_0({\bf x}_A)\right)}{e^0_0({\bf x}_A)} S_A^{ij}+\cdots,
\eeq
with $\kappa=1 $ and $1/2$ for the covariant SSC and  Newton-Wigner (NW) SSC respectively. To the order we work in this paper the vierbeins in Eq. (\ref{ssc1}) are given by\footnote{The effects on the dynamics (and on the preservation of the SSC) due to radiation do not enter until higher orders.}
\bea
e^0_0({\bf x}_A) &=& 1 - \frac{G m_B}{r} + \cdots \\ 
e^k_0({\bf x}_A) &=& - 2 \frac{G m_B}{r} {\bf v}_B^k + \frac{G}{r^2} ({\bf n}\times {\bf S}_B)^k +\cdots  \\ 
e^i_j({\bf x}_A) &=& \left(1+\frac{G m_B}{r}\right) \delta^i_j +\cdots
\eea
with ${\bf r} \equiv {\bf x}_A-{\bf x}_B$, $r \equiv |\mathbf r|$ and $\mathbf n \equiv \mathbf r / r$. 

Throughout the whole paper we impose the condition $B \neq A$ whenever we employ the index $B$ as a label for a binary constituent. With these we obtain 
\begin{align}
S_A^{i0} &= \kappa ~S_A^{ij}\left[{\bf v}_A^j+ \frac{G}{r}\left(2 m_B ({\bf v}_A^j-{\bf v}_B^j)+ \frac{1}{r^2} ({\bf r}\times {\bf S}_B)^j\right)+ (1-\kappa)\frac{{\bf v}_A^2}{2}{\bf v}_A^j \right] +\cdots \notag \\ 
&= \kappa \left[({\bf v}_A\times {\bf S}_A)^i \left(1+ (1-\kappa)\frac{{\bf v}_A^2}{2}\right)
+ \frac{2Gm_B}{r}\left(({\bf v}_A-{\bf v}_B)\times{\bf S}_A\right)^i+\frac{G}{r^3} \left({\bf S}_B^i{\bf r}\cdot {\bf S}_A-{\bf r}^i{\bf S}_A\cdot{\bf S}_B\right)\right] \notag \\ 
& +\cdots  \label{ssc}
\end{align}
for the on-shell relation between $S^{i0}_A$ and the spin 3-vector in the NW and covariant SSCs.

The alert reader may wonder at this point about the extra term in the Routhian of \cite{nrgrproc,nrgrss,nrgrs2}, schematically of the form $RSSuu$, needed to ensure the SSC is preserved in time. Even though this term vanishes on-shell, it does contribute to the equations of motion and to radiation. However, one can show that this extra piece does not enter in the multipole moments until 4PN, see Appendix \ref{app:B}.

\subsection{Finite Size Effects}

The LO spin(1)spin(1) contributions to the multipole moments arise due to finite size effects. Contrary to the case of conservative effects,  the ${\cal O}({\bf S}^2_A)$ contributions at NLO needed for the 3PN energy flux arise purely due to finite size operators. Here, we give an overview of the necessary Feynman rules for these terms, see \cite{nrgrs2} for further details.

\begin{figure}[t!]
    \centering
    \includegraphics[width=9cm]{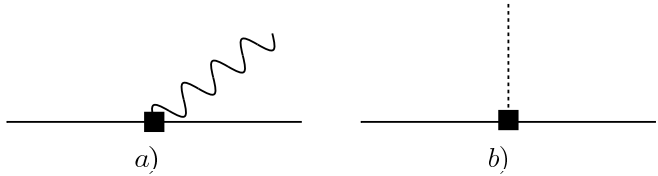}
\caption{\sl ${\cal O}({\bf S}_A^2)$ worldline coupling to: a) radiation and b) potential modes. A box represents couplings quadratic in spin due to the finite size operator in $L_{ES^2}$.}\label{worldc}
\end{figure}

In the EFT framework of NRGR finite size effects are naturally encoded by a set of higher dimensional operators in the worldline action which can be written in terms of the electric and magnetic components of the Weyl (or Riemann) tensor\footnote{Operators with Ricci tensor or scalar are redundant and can be removed via field redefinitions \cite{nrgr, Blanchet:2005tk}.} \cite{nrgr,dis1,nrgrs}. In addition, these operators absorb, via standard renormalization procedures, the divergences inherited from the point-particle approximation.  In treating spin effects, the first higher dimensional operator we encounter is the self-induced term
\cite{nrgrs,eih,nrgrs2} ($A=1,2$)
\begin{equation}
L_{ES^2} = \frac{C_{ES^2}^{(A)}}{2m_A}\frac{\tilde E_{ab}}{\sqrt{u^2}}
S_{Ac}^a S_A^{cb}\label{s2nc}, 
\end{equation}  
where $\tilde E_{ab} \equiv R_{\mu \alpha \nu \beta}v^{\mu} e^\alpha_a v^\nu e^\beta_b$ (the electric component of the Riemann tensor), and $C_{ES^2}^{(A)}$ are coefficients describing the short-distance physics at the scale $R$. For Kerr black holes $C_{ES^2}=1$, and the operator in Eq. (\ref{s2nc}) describes the self-induced mass quadrupole moment of a rotating black hole \cite{nrgrs,nrgrs2}. In the weak gravity approximation we have at linear order in $h_{\mu\nu}$ \cite{nrgrs2} 
\begin{eqnarray}
L_{ES^2}^{0} &=& -\frac{C_{ES^2}}{4m m_p}h_{00,ij}S^{ik}S^{jk},\label{fsr1}\\
L_{ES^2}^{v} &=& -\frac{C_{ES^2}}{2m m_p} h_{0l,ij}v^l S^{ik}S^{jk},\label{fsr2}\\
L_{ES^2}^{v^2} &=& \frac{C_{ES^2}}{2m
m_p}\left[\frac{1}{2}h_{00,ij}S^{i0}S^{j0}+
 S^{ik} S^{jk}\left(h_{il,0j}v^l-\frac{1}{2}h_{lr,ij}v^rv^l + h_{li,jr}v^lv^r +\frac{1}{2}h_{ij,l}a^l \right. \right. \nonumber \\ & & \left. \left. \hspace*{35pt}  -\frac{\mathbf{v}^2}{4}h_{00,ij}\right) + S^{0k} S^{jk}h_{00,lj}v^l\right]\label{fsr3},
\end{eqnarray}
with $a^l \equiv \frac{d v^l}{dt}$ the particle's acceleration, and the quadratic couplings we need are
\bea
L_{ES^2}^{h^2} &=& \frac{C_{ES^2}}{8mm^2_p} S^{ik} S^{jk} \Big[h_{00,i}h_{00,j}+ h_{00,l}(h_{ij,l}-2h_{il,j}) + h_{00}h_{00,ij} - 2 h_{li}h_{00,lj}\Big] + \ldots \label{fsh2}
\eea
Once again we split the gravitational field into potential and radiation modes, $h_{\mu\nu}=\bar h_{\mu\nu} + H_{\mu\nu}$, to generate the Feynman rules some of which are depicted in Figs. \ref{worldc}$ab$.

\subsection{Radiation with spin}

For the spin induced multipole moments the task is to compute all Feynman diagrams with an external radiated graviton and worldline spin couplings which contribute to the (spin-dependent) stress tensor $T^{\mu\nu}(t,{\bf k})$ in Eq. (\ref{Srad}) at the desired order. From there we obtain the multipole moments using Eqs.  (\ref{Q0ij} - \ref{Jijk}) and (\ref{fourier}). 

Some of these diagrams are such that  no potential modes are involved, and simply consist of a single radiation mode coupling to each worldline independently (depicted in Figs.~\ref{world}$a$ and \ref{worldc}$a$ for the spin and finite size couplings from Eqs. (\ref{sgnr1} - \ref{sgnr25}) and (\ref{fsr1} - \ref{fsr3}) respectively.) We refer to these contributions as `worldline radiation' since one can simply read off $T^{\mu \nu}$ from the worldline action. 

The remaining diagrams represent nonlinear gravitational effects and stem from two different sources: 1. The explicit nonlinear terms from the spin couplings such as $SH\bar h$ and $S^2 H\bar h$ in Eqs. (\ref{sgh2}) and (\ref{fsh2}). 2. The bulk action, which starts with the term $HH\bar h$, namely the `three-graviton' vertex\footnote{We obtain its Feynman rule by expanding the Einstein-Hilbert action plus gauge fixing term \cite{nrgr}.}. This is all we need in this paper. In these cases the radiation is emitted off the stress-energy of the system that includes the gravitational component.\\ 

Throughout our computations we encounter contributions linear and quadratic in spin. Components of cubic order in spin are omitted since they first contribute at 3.5PN. We extract the multipole moments from moments of $T^{\mu \nu}$ which in turn are computed from Feynman diagrams yielding $T^{\mu\nu}(t, \mathbf k)$. To clarify at which order spin effects start to conribute to the multipole moments, it is useful to know the scaling of the leading effects in $T^{\mu \nu}$. They follow from the linearized couplings in Eqs. (\ref{sgnr1}-\ref{sgnr2}) and (\ref{fsr1}-\ref{fsr3}) as well as their spinless counterparts. Since even the leading terms in $T^{\mu \nu}$ itself do not scale uniformly, we instead give the scalings for the leading contributions to general moments $K^{\mu \nu}_\ell \equiv \int d^3x T^{\mu \nu} \mathbf x^{i_1} \dots \mathbf x^{i_\ell}$ in Table \ref{tabl_Tscalings}. The scalings in Table \ref{tabl_Tscalings} also clarify which moments of $T^{\mu \nu}$ in the expressions for the multipole moments in Eqs. (\ref{Q0ij}-\ref{Jijk}) first contribute at a given order, where we remind the reader that time derivatives scale as $v/r$.
\begin{center}
 \begin{table}
  \begin{tabular}{| r | c | c | c |}
    \hline
    {} & ${\cal O}(\not {\hspace*{-2pt}\mathbf S})$ & ${\cal O}({\bf S}_A)$ & ${\cal O}({\bf S}_A^2)$\\ \hline
    $K^{00}_\ell$ & $m r^\ell$ & $m r^\ell v^3$ & $m r^\ell v^4$ \\ \hline
    $K^{0i}_\ell$ & $m r^\ell v$ & $m r^\ell v^2$ & $m r^\ell v^5$ \\ \hline    
    $K^{ij}_\ell$ & $m r^\ell v^2$ & $m r^\ell v^3$ & $m r^\ell v^6$ \\ \hline    
  \end{tabular}
  \caption{Scalings for leading terms in moments of $T^{\mu \nu}(t, \mathbf k)$ for various orders in spin. These scalings are valid for all moments $\ell \geq 2$.}
  \label{tabl_Tscalings}
 \end{table}
\end{center}

Below we review in more detail which are the the necessary ingredients for the energy flux to 3PN order.

\subsubsection{Spin-orbit radiation}

Spin-orbit effects in the energy flux at LO arise both from mass quadrupole and current quadrupole radiation, and give 1.5PN corrections to the LO spinless energy flux. 

More precisely, the leading ${\cal O}({\bf S}_A)$ effects in the energy flux (at 1.5PN) can enter in three different ways: 1. Due to the interference of the leading spinless mass quadrupole (0PN) with the leading ${\cal O}({\bf S}_A)$ mass quadrupole (1.5PN), 2. Due to the interference of the leading spinless current quadrupole (0PN) with the leading ${\cal O}({\bf S}_A)$ current quadrupole (0.5PN) or 3. Due to the spinless mass quadrupole radiation (0PN) but using the LO spin-orbit equations of motion (1.5PN).

Now for the ${\cal O}({\bf S}_A)$ effects in the energy flux at NLO or 2.5PN, the additional factor of $v^2$ may enter in many ways. However, it is clear that we need the mass quadrupole and current quadrupole, both spinless and ${\cal O}({\bf S}_A)$, to NLO, as well as the spinless and spin-orbit equations of motion also to NLO \cite{nrgr,nrgrso}. Moreover, we have to include the LO mass octupole and the LO current octupole, both spinless and ${\cal O}({\bf S}_A)$. As we shall see, the matching calculation of the mass quadrupole and current quadrupole at NLO requires the inclusion of nonlinear effects.

Since the leading ${\cal O}({\bf S}_A)$ radiation effects arise at 1.5PN, and the leading tail effect  yields a 1.5PN correction to the flux, we also need to include the tail contribution to the energy flux at 3PN linear in spin, see \cite{rad1} for a generic expression to compute such effects.

\subsubsection{Spin-spin radiation}

The LO spin-spin contributions to the energy flux first appear at 2PN. They can for example arise when squaring the leading ${\cal O}({\bf S}_A)$ current quadrupole, which gives both ${\cal O}({\bf S}_A {\bf S}_B)$ and ${\cal O}({\bf S}_A^2)$ terms in the energy flux. Note that the leading ${\cal O}({\bf S}_A)$ mass quadrupole (1.5PN) does not contribute to spin-spin effects in the energy flux until 3PN. Therefore, the only ${\cal O}({\bf S}_A)$ multipole moment needed to NLO for the spin-spin effects in the 3PN energy flux is the current quadrupole, whereas the leading ${\cal O}({\bf S}_A)$ expressions for the mass quadrupole and the current octupole are sufficient.

Besides the ${\cal O}({\bf S}_A)$ contributions to the multipole moments discussed above we also have ${\cal O}({\bf S}_A^2)$ as well as ${\cal O}({\bf S}_A {\bf S}_B)$ components. The LO contribution to the ${\cal O}({\bf S}_A^2)$ energy flux (at 2PN) from the ${\cal O}({\bf S}_A^2)$ multipole moments arises due to the leading ${\cal O}({\bf S}_A^2)$ contribution to the mass quadrupole (2PN). For the ${\cal O}({\bf S}_A^2)$ power at 3PN we compute the ${\cal O}({\bf S}_A^2)$ mass quadrupole to NLO as well as the leading current quadrupole and the leading mass octupole at ${\cal O}({\bf S}_A^2)$, along with their spinless expressions. 

In the EFT that lives between the scales $R$ and $r$ there cannot be any worldline coupling of ${\cal O}({\bf S}_A {\bf S}_B)$. Thus, ${\cal O}({\bf S}_A {\bf S}_B)$ worldline radiation cannot exist and the Feynman diagrams for every ${\cal O}({\bf S}_A {\bf S}_B)$ component of the multipole moments must involve potential graviton lines connecting the two worldlines, where on each worldline there has to be a coupling linear in spin. The leading contribution of this sort arises for the mass quadrupole moment at 3PN. For the power at 3PN, we only need to calculate this leading ${\cal O}({\bf S}_A {\bf S}_B)$ component for the mass quadrupole. However, there is another 3PN contribution to the mass quadrupole at ${\cal O}({\bf S}_A {\bf S}_B)$ which arises indirectly due to the replacement of $S^{i0}$ using Eq. (\ref{ssc}) in the leading mass quadrupole at ${\cal O}({\bf S}_A$). This is similar to what happened for the NLO conservative spin(1)spin(2) dynamics where the leading order spin-orbit diagrams contributed a term through the replacement of $S_A^{i0}$ \cite{comment, nrgrss}.\\

Furthermore, we recall that the spin-dependence in the energy flux to 3PN may always enter from spinless multiple moments by using the spin-dependent conservative equations of motion that derive from the potentials in \cite{nrgr,nrgrs,nrgrso,nrgrss,nrgrs2} to NLO.

\section{Spin contributions to the source multipole moments}\label{sec:match}

Now we set out to compute the necessary Feynman diagrams which yield all spin contributions to the source multipole moments for the energy flux to 3PN. The spin-independent multipoles which enter the spin-dependent part of the energy flux to 3PN are calculated partly in \cite{rad1} and in Appendix \ref{app:C}. 

\subsection{Mass quadrupole moment $I^{ij}$}

We begin with the calculation of the mass quadrupole which is the most elaborate one. We present first the contributions from worldline radiation which arises from diagrams without any potential modes and then calculate the more complicated diagrams involving potential modes and nonlinearities.

\subsubsection{${\cal O}({\bf S}_A)$ worldline radiation}

The Feynman diagram which gives rise to worldline radiation linear in spin is depicted in Fig.~\ref{world}$a$.
To compute these contributions we can either consider $T^{\mu \nu}$ in coordinate space or in mixed coordinate-momentum space. From Eq. (\ref{act2}) we obtain the compact expression for the contribution to $T^{\mu \nu}$ from Fig.~\ref{world}$a$ in coordinate space \cite{nrgrs}
\beq
\label{set}
T_{\ref{world}a}^{\mu\nu} (x) = \frac{1}{2} \sum_A \int dt' ~\partial_\alpha \delta^4(x -x_A(t')) \left(S_A^{\nu\alpha}(t') u_A^\mu(t')+ S_A^{\mu\alpha}(t') u_A^\nu(t')\right),
\eeq 
where $x \equiv (t,{\bf x})$, $\partial_\alpha \equiv \frac{\partial}{\partial x^\alpha}$ and $x_A(t')\equiv (t',{\bf x}_A(t'))$. From now on we will suppress the implicit time dependence of the worldline variables.
The expressions in mixed coordinate-momentum space read
\begin{align}
 T_{\ref{world}a}^{00}(t, \mathbf k) & = \sum_A S_A^{0i} \, i \mathbf k^i e^{-i {\mathbf k} \cdot \mathbf x_A} \label{eq:WLrad00SA}\\
 T_{\ref{world}a}^{0i}(t, \mathbf k) & = \frac{1}{2} \sum_A \left(S_A^{ij} \, i \mathbf k^j + S_A^{0j} \mathbf v_A^i \, i \mathbf k^j + S_A^{0i} \, i \mathbf k \cdot \mathbf v_A - \dot S^{0i} \right) e^{-i {\mathbf k} \cdot \mathbf x_A} \label{eq:WLrad0iSA}\\
 T_{\ref{world}a}^{ij}(t, \mathbf k) & = \frac{1}{2}\sum_A \left\{\left(S_A^{il} \mathbf v_A^j + S_A^{jl} \mathbf v_A^i \right) i \mathbf k^l + i \mathbf k \cdot \mathbf v_A \left(S_A^{0i} \mathbf v_A^j + S_A^{0j} \mathbf v_A^i \right) \right. \notag \\
& \left. {} \hspace*{39pt} - \dot S_A^{0i} \mathbf v_A^j - \dot S_A^{0j} \mathbf v_A^i - S_A^{0i} \mathbf a_A^j - S_A^{0j} \mathbf a_A^i \right\} e^{-i {\mathbf k} \cdot \mathbf x_A} \label{eq:WLradijSA}.
\end{align} 
For the ${\cal O}({\bf S}_A)$ mass quadrupole at NLO the subleading terms in the components of $T^{\mu \nu}$ are required in the pieces of $I^{ij}$ of Eqs. (\ref{Q0ij} - \ref{Q2ij}) which give the leading ${\cal O}({\bf S}_A)$ mass quadrupole,
\begin{equation}
 \int d^3{\bf x} \left( T^{00} + T^{ll}-\frac{4}{3}\dot T^{0l}{\bf x}^l \right) [{\bf x}^i{\bf x}^j]_{\rm STF} \label{eq:IijSALO},
\end{equation}
whereas the higher order terms
\begin{equation}
 \int d^3{\bf x} \left(\frac{11}{42} \ddot T^{00}{\bf x}^2 + \frac{2}{21} \ddot T^{ll} {\bf x}^2 + \frac{1}{6} \ddot T^{lm} {\bf x}^l {\bf x}^m - \frac{1}{7} \dddot T^{0l} {\bf x}^l {\bf x}^2\right) [{\bf x}^i{\bf x}^j]_{\rm STF} \label{eq:IijSANLO}
\end{equation}
only require the leading terms in the components of $T^{\mu \nu}$.

Using either form of $T^{\mu \nu}$ it is straightforward to compute the moments in Eqs. (\ref{eq:IijSALO}, \ref{eq:IijSANLO}). Up to NLO we get for the mass quadrupole contributions from Fig.~\ref{world}$a$
\bea
\label{Qijsglin}
 I_{\ref{world}a}^{ij} & = & \sum_A \left[\left(2\kappa+\frac{2}{3}\right) ({\bf v}_A\times {\bf S}_A)^i {\bf x}_A^j  -\frac{4}{3} ({\bf x}_A\times {\bf S}_A)^i{\bf v}_A^j-\frac{4}{3}({\bf x}_A\times \dot {\bf S}_A)^i{\bf x}_A^j \right. \nn \\
& & \left. \hspace*{18pt} - \frac{d}{dt} \left\{\frac{4\kappa}{3}{\bf v}_A \cdot {\bf x}_A({\bf v}_A\times{\bf S}_A)^i{\bf x}_A^j\right\} \right. \nn \\
& & \left. \hspace*{18pt} +\frac{d^2}{dt^2}\left\{\frac{1+11\kappa}{21} {\bf x}_A^2 ({\bf v}_A\times{\bf S}_A)^i{\bf x}_A^j + \frac{1}{7}{\bf x}_A^2 ({\bf S}_A\times{\bf x}_A)^i{\bf v}_A^j \right. \right. \nn \\ & & 
\left. \left. {} \hspace*{45pt} +  \frac{1}{21} {\bf x}_A\cdot{\bf v}_A  ({\bf x}_A\times{\bf S}_A)^i{\bf x}_A^j+\frac{1-6\kappa}{42}({\bf v}_A\times{\bf S}_A)\cdot {\bf x}_A~{\bf x}_A^i{\bf x}_A^j \right\} \right]_{\rm STF} \nn\\ & + &  \kappa \sum_{A, B}\frac{2G}{r} \left[ 2m_B \left(({\bf v}_A-{\bf v}_B)\times{\bf S}_A\right)^i{\bf x}_A^j  + \frac{1}{r^2}\left({\bf S}_A\cdot {\bf r} {\bf S}_B^i {\bf x}_A^j-{\bf S}_A\cdot {\bf S}_B {\bf r}^i{\bf x}_A^j\right) \right]_{\rm STF}\nn \\
& + &  2 \kappa (1-\kappa)\sum_A \left[\frac{{\bf v}^2_A}{2} \left({\bf v}_A\times{\bf S}_A\right)^i{\bf x}_A^j\right]_{\rm STF},
\eea
where we applied the replacement rule for $S_A^{i0}$ in Eq. (\ref{ssc}) and used $S_A^{ij} = \epsilon^{ijk} \mathbf S_A^k$. The corrections due to Eq. (\ref{ssc}) are the terms in the last two lines of Eq. (\ref{Qijsglin}), and we notice that this replacement also induces an ${\cal O}({\bf S}_A {\bf S}_B)$ contribution to the mass quadrupole which is of order 3PN and which we have to keep. Furthermore, note that we kept a term proportional to $\dot {\bf S}_A$. Upon use of the spin equations of motion (which we will not perform until we compute the energy flux in \cite{powerloss}) this term becomes an ${\cal O}({\bf S}_A)$ contribution at 2.5PN and spin-spin contributions at 3PN to the mass quadrupole moment. Employing the covariant SSC $\kappa = 1$, Eq. (\ref{Qijsglin}) becomes
\bea
 I_{\ref{world}a}^{ij} & = & \sum_A \left[\frac{8}{3} ({\bf v}_A\times {\bf S}_A)^i {\bf x}_A^j  -\frac{4}{3} ({\bf x}_A\times {\bf S}_A)^i{\bf v}_A^j-\frac{4}{3}({\bf x}_A\times \dot {\bf S}_A)^i{\bf x}_A^j -  \frac{4}{3} \frac{d}{dt} \left\{{\bf v}_A \cdot {\bf x}_A({\bf v}_A\times{\bf S}_A)^i{\bf x}_A^j\right\} \right. \nn \\
& & \left. {} \hspace*{18pt} +\frac{1}{7}\frac{d^2}{dt^2}\left\{ 4{\bf x}_A^2 ({\bf v}_A\times{\bf S}_A)^i{\bf x}_A^j+ \frac{1}{3} {\bf x}_A\cdot{\bf v}_A  ({\bf x}_A\times{\bf S}_A)^i{\bf x}_A^j  \right. \right. \nn \\ 
& & \left. \left. {} \hspace*{57pt} +  {\bf x}_A^2 ({\bf S}_A\times{\bf x}_A)^i{\bf v}_A^j-\frac{5}{6}({\bf v}_A\times{\bf S}_A)\cdot {\bf x}_A~{\bf x}_A^i{\bf x}_A^j \right\} \right]_{\rm STF} \nn  \\ 
& + & \sum_{A, B}\frac{2G}{r} \left[ 2m_B \left(({\bf v}_A-{\bf v}_B)\times {\bf S}_A\right)^i {\bf x}_A^j  + \frac{1}{r^2}\left({\bf S}_A\cdot {\bf r} {\bf S}_B^i {\bf x}_A^j-{\bf S}_A\cdot {\bf S}_B {\bf r}^i{\bf x}_A^j\right) \right]_{\rm STF}.
\label{Q012ijnlo}
\eea

\subsubsection{${\cal O}({\bf S}_A^2)$ worldline radiation}

The Feynman diagram for ${\cal O}({\bf S}_A^2)$ worldline radiation is shown in Fig.~\ref{worldc}$a$.  The relevant couplings arise due to Eqs. (\ref{fsr1} - \ref{fsr3}). The expressions for $T^{\mu \nu}$ from Fig.~\ref{worldc}$a$ are
\begin{align}
  T^{00}_{\ref{worldc}a}(t, \mathbf k) & = - \sum_A \frac{C_{ES^2}^{(A)}}{2 m_A} \left[\left(1 + \frac{\mathbf v_A^2}{2}\right) S_A^{ik} S_A^{jk} - S_A^{0i} S_A^{0j} - 2 S_A^{ik} S_A^{0k} \mathbf v_A^j\right] \mathbf k^i \mathbf k^j e^{-i {\mathbf k} \cdot \mathbf x_A}\\ 
  T^{0i}_{\ref{worldc}a}(t, \mathbf k) & = - \sum_A \frac{C_{ES^2}^{(A)}}{2 m_A} S_A^{jl} S_A^{kl} \mathbf v_A^i \, \mathbf k^j \mathbf k^k e^{-i {\mathbf k} \cdot \mathbf x_A}  \label{eq:T0iSAsq} \\
  T^{ll}_{\ref{worldc}a}(t, \mathbf k) & = - \sum_A \frac{C_{ES^2}^{(A)}}{2 m_A} \left[- S_A^{kl} S_A^{kl} \, i \mathbf k \cdot \mathbf a_A + S_A^{kl} S_A^{ml} \mathbf v_A^2 \mathbf k^k \mathbf k^m + 2 S_A^{km} S_A^{lm} \mathbf a_A^l i \mathbf k^k \right] e^{-i {\mathbf k} \cdot \mathbf x_A} 
\end{align}
where we dropped time derivatives of the spin variables which do not contribute at the order we are working. From here we obtain the moments 
{ \allowdisplaybreaks
\begin{align}
 \int d^3 {\bf x} T_{\ref{worldc}a}^{00} \left[{\bf x}^i {\bf x}^j\right]_{\rm STF} & = \sum_A \frac{C_{ES^2}^{(A)}}{2 m_A} \left[\left(-2 + (2 \kappa^2 - 1) \mathbf v_A^2 \right) \mathbf S_A^i \mathbf S_A^j - 4 \kappa (\kappa - 1) \mathbf S_A \cdot \mathbf v_A \mathbf S_A^i \mathbf v_A^j \right. \notag \\
& {} \hspace*{59pt} \left. + 2 \kappa (\kappa - 2) \mathbf S_A^2 \mathbf v_A^i \mathbf v_A^j\right]_{\rm STF} \\
 \int d^3 {\bf x} T_{\ref{worldc}a}^{ll} \left[{\bf x}^i {\bf x}^j\right]_{\rm STF} & = \sum_A \frac{C_{ES^2}^{(A)}}{m_A} \left[- \mathbf v_A^2 \mathbf S_A^i \mathbf S_A^j - 2 \mathbf S_A \cdot \mathbf a_A \mathbf S_A^i \mathbf x_A^j \right]_{\rm STF} \\
 \int d^3 {\bf x} \dot T_{\ref{worldc}a}^{0l} {\bf x}^l \left[{\bf x}^i {\bf x}^j\right]_{\rm STF} & = \frac{d}{dt} \sum_A \frac{C_{ES^2}^{(A)}}{m_A} \left[\mathbf x_A \cdot \mathbf v_A S_A^{ik} S_A^{jk} + 2 S_A^{kl} \mathbf v_A^k S_A^{il} \mathbf x_A^j \right]_{\rm STF} \notag \\
      & = \sum_A \frac{C_{ES^2}^{(A)}}{m_A} \left[- (\mathbf v_A^2 + \mathbf x_A \cdot \mathbf a_A) \mathbf S_A^i \mathbf S_A^j + 2 \mathbf S_A^2 (\mathbf  v_A^i \mathbf  v_A^j + \mathbf  x_A^i \mathbf a_A^j) \right. \notag \\
      & {} \hspace*{63pt} \left. - 2 \mathbf S_A \cdot \mathbf v_A \mathbf S_A^i \mathbf v_A^j - 2 \mathbf S_A \cdot \mathbf a_A \mathbf S_A^i \mathbf x_A^j \right]_{\rm STF} \\
 \int d^3 {\bf x} \ddot T_{\ref{worldc}a}^{00} \mathbf x^2 \left[{\bf x}^i {\bf x}^j\right]_{\rm STF} & = \frac{d^2}{dt^2} \sum_A \frac{C_{ES^2}^{(A)}}{m_A} \left[\mathbf x_A^2 S_A^{ik} S_A^{jk} + 4 S_A^{kl} x_A^k S_A^{il} \mathbf x_A^j + S_A^{kl} S_A^{kl} \mathbf x_A^i \mathbf x_A^j \right]_{\rm STF} \notag \\
      & = \sum_A \frac{C_{ES^2}^{(A)}}{m_A} \left[- 2 (\mathbf v_A^2 + \mathbf x_A \cdot \mathbf a_A) \mathbf S_A^i \mathbf S_A^j + 12 \mathbf S_A^2 (\mathbf  v_A^i \mathbf  v_A^j + \mathbf  x_A^i \mathbf a_A^j) \right. \notag \\
      & {} \hspace*{63pt} \left. - 8 \mathbf S_A \cdot \mathbf v_A \mathbf S_A^i \mathbf v_A^j - 4 \mathbf S_A \cdot \mathbf a_A \mathbf S_A^i \mathbf x_A^j - 4 \mathbf S_A \cdot \mathbf x_A \mathbf S_A^i \mathbf a_A^j \right]_{\rm STF} 
\end{align} }
which we combine to the mass quadrupole using Eqs. (\ref{Q0ij}, \ref{Q1ij}). We obtain
\bea
I^{ij}_{\ref{worldc}a} &=&  \sum_A \frac{C_{ES^2}^{(A)}}{m_A}\left[{\bf S}_A^j{\bf S}_A^i\left(-1+{\bf v}_A^2\left(\kappa^2-\frac{29}{42}\right)+\frac{17}{21} {\bf a}_A\cdot {\bf x}_A\right)-\frac{8}{21}{\bf x}^i_A{\bf S}_A^j {\bf a}_A\cdot {\bf S}_A\nonumber\right.\\ & & \left. {} \hspace*{46pt} -\frac{22}{21}{\bf a}_A^i{\bf S}_A^j {\bf S}_A\cdot {\bf x}_A+\left(\frac{4}{7}-2(\kappa-1)\kappa\right){\bf v}^i_A{\bf S}_A^j{\bf S}_A\cdot {\bf v}_A \right. \nn \\ & & \left. {} \hspace*{46pt} +{\bf S}_A^2\left(\left(\kappa(\kappa-2)+\frac{10}{21}\right){\bf v}^i_A{\bf v}^j_A + \frac{10}{21} {\bf a}_A^i{\bf x}_A^j\right) \right]_{\rm STF}
\eea
and in the covariant SSC we have
\bea
\label{Qijfslin}
I^{ij}_{\ref{worldc}a}&=& \sum_A \frac{C_{ES^2}^{(A)}}{m_A}\left[{\bf S}_A^i{\bf S}_A^j\left(-1+\frac{13}{42} {\bf v}_A^2+\frac{17}{21} {\bf a}_A\cdot {\bf x}_A\right)+{\bf S}_A^2\left(-\frac{11}{21}{\bf v}^i_A{\bf v}^j_A +\frac{10}{21}  {\bf a}_A^i{\bf x}_A^j\right)\right.\nonumber\\ & & \left.  {} \hspace*{46pt} -  \frac{8}{21}{\bf x}^i_A{\bf S}_A^j {\bf a}_A\cdot {\bf S}_A+\frac{4}{7}{\bf v}^i_A{\bf S}_A^j{\bf S}_A\cdot {\bf v}_A-\frac{22}{21}{\bf a}_A^i{\bf S}_A^j {\bf S}_A\cdot {\bf x}_A
\right]_{\rm STF}.
\eea

\subsubsection{Nonlinear gravitational contributions at ${\cal O}({\bf S}_A {\bf S}_B)$}

\begin{figure}[h!]
    \centering
    \includegraphics[width=11cm]{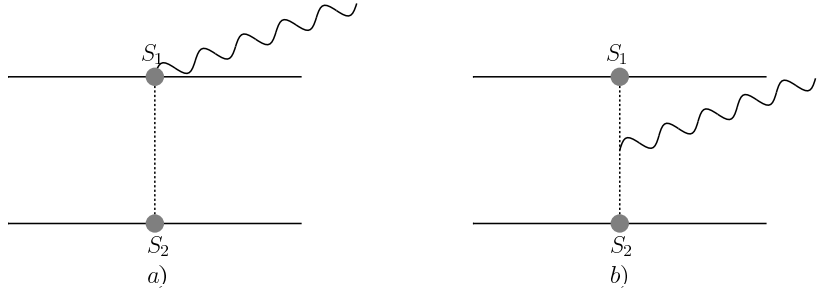}
\caption{\sl Nonlinear gravitational effects at ${\cal O}({\bf S}_1 {\bf S}_2)$.} \label{nonlSS}
\end{figure}

Let us start the discussion of nonlinear diagrams with the new results quadratic in spin and first consider the spin(1)spin(2) contributions.
For the leading ${\cal O}({\bf S}_A {\bf S}_B)$ mass quadrupole moment (which is of order 3PN) we need to calculate Feynman diagrams which involve the exchange of a potential graviton shown in Figs. \ref{nonlSS}$ab$. 

The radiation graviton can then either couple to the worldline together with a potential graviton (Fig. \ref{nonlSS}$a$) or to the potential graviton away from the worldline (Fig. \ref{nonlSS}$b$).\\

Let us calculate in somewhat more detail the contribution to $I^{ij}$ from the diagram in Fig.~\ref{nonlSS}$a$, beginning with $T^{00}$. For this we need the $H S \bar h_{00}$ coupling in Eq. (\ref{sgh2}), concretely $\frac{1}{4m_p^2} S_1^{li} H_{0i}\bar h_{00,l}$, that we ought to contract with the LO spin vertex in Eq. (\ref{sgnr1}). Then the contribution to the effective action in Eq. (\ref{Srad}) will be of the form (in momentum space)
\beq
-\frac{i}{2m_p} \int d^3{\bf k} T^{00}_{\ref{nonlSS}a}(t,{\bf k}) h_{00}(t,-{\bf k})  = \frac{i}{4m_p^2}\frac{i}{2m_p}S_1^{li} S_2^{km}\langle H_{0i}({\bf x}_1)H_{0k,m}({\bf x}_2)\rangle \bar h_{00,l}({\bf x}_1).
\eeq

Using 
\beq
\langle H_{0i}({\bf x}_A)H_{0k,l}({\bf x}_B)\rangle = P_{0i0k} \int \frac{d^3 {\bf q}}{(2\pi)^3} \frac{-i}{ {\bf q}^2 }~(i{\bf q}^l)~
e^{-i {\bf q}\cdot ({\bf x}_A-{\bf x}_B)} = i\delta^{ik}\frac{1}{8\pi} \frac{{\bf n}^l}{r^2}, 
\eeq
we have (after summing over mirror images)
\beq
T^{00}_{\ref{nonlSS}a}({\bf k})= \sum_{A, B} \frac{G}{r^3} \left(({\bf r}\times {\bf S}_B)\times {\bf S}_A\right)^l (-i{\bf k}^l) e^{-i{\bf k}\cdot {\bf x}_A}.
\eeq

Similarly, for $T^{lk}$ we need the coupling $\frac{1}{4m_p^2}S^{il}H_{0k,i}\bar h_{lk}$ from Eq. (\ref{sgh2}). Following the same steps the result is (after tracing over $lk$)
\beq
T^{ll}_{\ref{nonlSS}a}({\bf k})=-\frac{2G}{r^3} \sum_{A, B} \left( {\bf S}_A\cdot {\bf S}_B - 3 {\bf S}_A\cdot {\bf  n} {\bf S}_B \cdot {\bf n}\right)  e^{-i{\bf k}\cdot {\bf x}_A}.
\eeq

Since the second and third terms in Eq. (\ref{Q1ij}) are clearly subleading, to compute the contribution to the mass quadrupole we just need to expand $T^{00}_{\ref{nonlSS}a}({\bf k})+T^{ll}_{\ref{nonlSS}a}({\bf k})$ to second order in ${\bf k}$ and read off the contribution to the mass quadrupole using (\ref{fourier}). The result is
\beq
 I^{ij}_{\ref{nonlSS}a}=  \sum_{A, B} \frac{2G}{r^3}\left[{\bf S}_A\cdot {\bf r} {\bf S}_B^i {\bf x}_A^j - {\bf S}_B\cdot {\bf S}_A {\bf r}^i{\bf x}_A^j -\left( {\bf S}_A\cdot {\bf S}_B - 3 {\bf S}_A\cdot {\bf n} {\bf S}_B \cdot {\bf n}\right) {\bf x}_A^i{\bf x}_A^j \right]_{\rm STF}.
\eeq 

For the diagram in Fig. \ref{nonlSS}$b$ we need the three-graviton coupling as well as two LO spin insertions of Eq. (\ref{sgnr1}). This diagram is more symmetric, and since we only need $T^{00}+T^{ii}$, we can combine both Feynman rules, i.e. we add the one with an external $\bar h_{00}$ and the one with an external ${\bar h}_{lk}$ which we trace over $kl$. The expression for the combined three-graviton vertex contracted with the two potential propagators reads
\beq
\langle H_{0i}({\bf p+k})H_{0j}({\bf p})[HH\bar h({\bf k})] \rangle = \frac{i}{4m_p}[2{\bf k}^i{\bf k}^j+2({\bf p}^i{\bf k}^j-{\bf k}^i{\bf p}^j)]\frac{-i}{{\bf p}^2}\frac{-i}{({\bf k+p})^2},
\eeq
where $[HH\bar h({\bf k})]$ represents the three-graviton vertex \cite{nrgr}. The open indices are contracted with the spin couplings on each worldline, then we have
\beq
(T^{00}+T^{kk})_{\ref{nonlSS}b}({\bf k})=\int \frac{d^3 {\bf p}}{(2\pi)^3}  \frac{1}{4m^2_p}[{\bf k}^i{\bf k}^j+({\bf p}^i{\bf k}^j-{\bf k}^i{\bf p}^j)] \frac{1}{{\bf p}^2}\frac{1}{({\bf k+p})^2} S_1^{ik} {\bf p}^k S_2^{jl}(-{\bf p}^l-{\bf k}^l) e^{i{\bf p}\cdot{\bf x}_1} e^{-i({\bf p+k})\cdot {\bf x}_2}.
\eeq

Expanding to to second order in momentum and after some cancelations we have
\beq
(T^{00}+T^{kk})_{\ref{nonlSS}b}({\bf k})=-\int \frac{d^3 {\bf p}}{(2\pi)^3} \frac{1}{2m^2_p}S_1^{ik}S_2^{jl}\left[\frac{{\bf k}^i{\bf k}^j{\bf p}^k{\bf p}^l}{{\bf p}^4} e^{i{\bf p} \cdot {\bf x}_1} e^{-i{\bf p}\cdot {\bf x}_2}\right] + \ldots, 
\eeq
and performing the Fourier integration we obtain
 \beq
(T^{00}+T^{kk})_{\ref{nonlSS}b}({\bf k})=\frac{4G}{r^3} S_1^{im}S_2^{jl}\left(r^2\delta^{ml}-{\bf r}^m{\bf r}^l \right) \left(-\frac{{\bf k}^i{\bf k}^j}{2}\right) + \cdots .
\eeq

Therefore from Eq. (\ref{fourier}) 
\beq
 I^{ij}_{\ref{nonlSS}b}=\sum_{A, B} \frac{2G}{r^3} \left[ {\bf r}^i{\bf r}^j {\bf S}_A\cdot {\bf S}_B -2{\bf S}_A\cdot {\bf r} {\bf S}_B^i{\bf r}^j\right]_{\rm STF}.
\eeq

The combined result from these diagrams reads
\beq
\label{3ab}
I^{ij}_{\ref{nonlSS}ab}=  \sum_{A, B} \frac{G}{2r^3}\left[
12{\bf x}_A^i{\bf x}_A^j{\bf S}_A\cdot {\bf n} {\bf S}_B\cdot {\bf n} - 4{\bf S}_A\cdot {\bf S}_B {\bf x}_A^i{\bf x}_B^j +  4{\bf S}_A\cdot {\bf r} (2{\bf S}_B^i{\bf x}_B^j-{\bf x}_A^i{\bf S}_B^j)\right]_{\rm STF}. 
\eeq

\subsubsection{Nonlinear gravitational contributions at ${\cal O}({\bf S}_A^2)$}

\begin{figure}[h!]
    \centering
    \includegraphics[width=9cm]{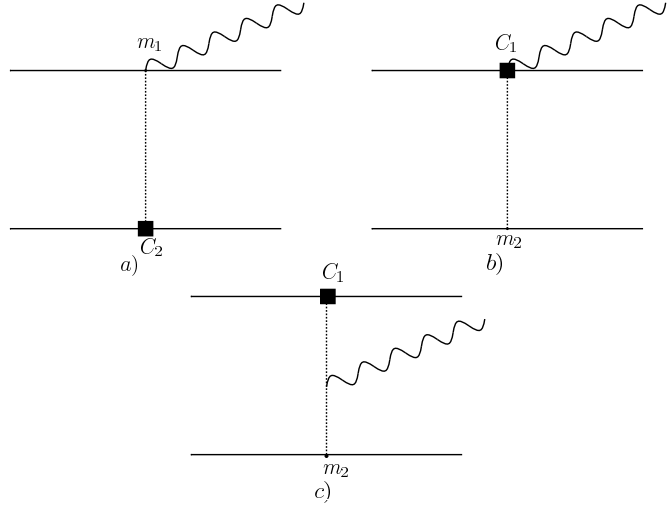}
\caption{\sl Nonlinear gravitational effects from finite size at ${\cal O}({\bf S}_A^2)$.} \label{nonlS2}
\end{figure}

We now calculate the finite size contributions. Even though the leading ${\cal O}({\bf S}_A^2)$ mass quadrupole is entirely due to $T^{00}$ worldline radiation, we need to consider nonlinear contributions not only to $T^{00}$ but also to $T^{ij}$. This is because at ${\cal O}({\bf S}_A^2)$, the leading $T^{ij}$ is not purely due to worldline radiation but gets a contribution from diagrams with potential exchange as well. This is the same as what happens in the spinless case \cite{rad1}. The three Feynman diagrams with potential exchange which enter the ${\cal O}({\bf S}_A^2)$ mass quadrupole at NLO are shown in Fig.~\ref{nonlS2}$abc$.

The calculations of the Feynman diagrams proceed analogously to the nonlinear ones at ${\cal O}({\bf S}_A {\bf S}_B)$ above. Notice that the first diagram in Fig.~\ref{nonlS2}$a$ only contributes to $T^{00}$ since the worldline coupling to $\bar h_{ij}$ introduces an extra power of $v^2$ making it a higher order term in $T^{ij}$.
We find
\beq
 \int d^3{\bf x} T^{00}_{\ref{nonlS2}a} [{\bf x}^i{\bf x}^j]_{\rm TF} = \sum_{A, B} \frac{m_AC_{ES^2}^{(B)}}{m_B}\frac{G}{2r^3}\left( {\bf S}_B^2 -3({\bf S}_B\cdot {\bf n})^2\right) [{\bf x}_A^i{\bf x}_A^j]_{\rm TF} 
\eeq
\beq
 \int d^3{\bf x} T^{00}_{\ref{nonlS2}b} [{\bf x}^i{\bf x}^j]_{\rm TF}= \sum_{A, B}\frac{GC_{ES^2}^{(B)}m_A}{2m_Br^3}\left[ 4 {\bf S}_B^2 {\bf x}_A^i{\bf x}_B^j+ 2r^2 {\bf S}_B^i{\bf S}_B^j-3\left({\bf S}_B^2+({\bf S}_B\cdot {\bf n})^2\right){\bf x}_B^i{\bf x}_B^j \right]_{\rm STF}\eeq
\beq
 \int d^3{\bf x} T^{kk}_{\ref{nonlS2}b} [{\bf x}^i{\bf x}^j]_{\rm TF}= \sum_{A, B}\frac{GC_{ES^2}^{(B)}m_A}{2m_Br^3} \left[4 {\bf S}_B\cdot {\bf r} {\bf S}_B^i{\bf x}_B^j -2\left({\bf S}_B^2-3({\bf S}_B\cdot {\bf n})^2\right) {\bf x}_B^i{\bf x}_B^j\right]_{\rm STF}. \eeq 
and 
\bea
I^{ij}_{\ref{nonlS2}c}=\int d^3{\bf x} (T^{00}+T^{kk})_{\ref{nonlS2}c} [{\bf x}^i{\bf x}^j]_{\rm TF} &=&  \sum_{A, B} \frac{GC_{ES^2}^{(B)} m_A}{2m_B r^3}\left[-2({\bf S}_B^2-3({\bf S}_B\cdot {\bf n})^2) ({\bf x}_B^i{\bf x}_B^j+{\bf x}_A^i{\bf x}_A^j) \right. \nn \\ & & \left. +4 r^2 {\bf S}_B^i{\bf S}_B^j - 8 {\bf S}_B^2 {\bf r}^i{\bf x}_B^j+8 {\bf S}_B\cdot {\bf r} {\bf S}_B^i{\bf x}_B^j\right]_{\rm STF} .
\eea
For the total contribution to the ${\cal O}({\bf S}_A^2)$ mass quadrupole from Figs.~\ref{nonlS2}$abc$ we have
\bea
I^{ij}_{\ref{nonlS2}abc}&=& \sum_{A, B} \frac{GC_{ES^2}^{(B)} m_A}{2m_B r^3}\left[({\bf S}_B^2+9({\bf S}_B\cdot {\bf n})^2){\bf x}_B^i{\bf x}_B^j+\left(3({\bf S}_B\cdot {\bf n})^2-{\bf S}_B^2\right){\bf x}_A^i{\bf x}_A^j \right. \nonumber \\ & & \left. +6 r^2 {\bf S}_B^i{\bf S}_B^j - 4 {\bf S}_B^2 {\bf x}_A^i{\bf x}_B^j+12 {\bf S}_B\cdot {\bf r} {\bf S}_B^i{\bf x}_B^j\right]_{\rm STF} \label{Qij4abc}.
\eea

\subsubsection{Nonlinear gravitational contributions at ${\cal O}({\bf S}_A)$}

\begin{figure}[h!]
    \centering
    \includegraphics[width=9cm]{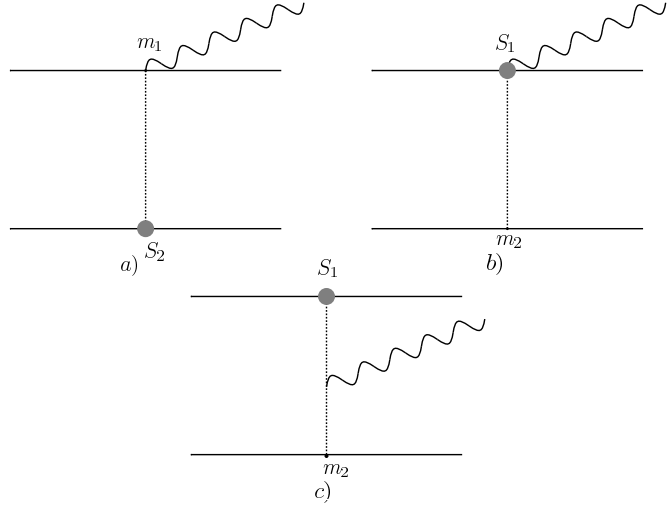}
\caption{\sl Nonlinear gravitational effects at ${\cal O}({\bf S}_A)$.} \label{nonlS}
\end{figure}

Finally for the nonlinear ${\cal O}({\bf S}_A)$ contributions there exist three Feynman diagrams as seen in Fig.~\ref{nonlS}$abc$.

For the NLO mass quadrupole at ${\cal O}({\bf S}_A)$ we also have to calculate subleading terms for the moments in Eq. (\ref{eq:IijSALO}), i.e. for all components of $T^{\mu \nu}$. This means that we need to consider the diagrams in Fig.~\ref{nonlS} for external radiation field legs with $\bar h_{00},\bar h_{0i}$ and $\bar h_{ij}$. The diagram in Fig. \ref{nonlS}$a$ only contributes at the order we are working for $\bar h_{00}$ since the other components introduce an additional power of $v^2$ in the couplings to the worldlines, hence
\beq
I^{ij}_{\ref{nonlS}a} = \int d^3 {\bf x}~T_{\ref{nonlS}a}^{00} [{\bf x}^i {\bf x}^j]_{\rm TF} = -\sum_{A, B} \frac{Gm_B}{r^3} \left\{(1+\kappa)({\bf v}_A\times {\bf S}_A)\cdot {\bf r}-2({\bf v}_B\times {\bf S}_A)\cdot {\bf r}\right\} \left[{\bf x}_B^i{\bf x}_B^j \right]_{\rm TF}. \eeq

For the contributions from the diagram in Fig. \ref{nonlS}$b$ we have
\bea
\int d^3 {\bf x}\left(T^{00}+T^{ii}\right)_{\ref{nonlS}b} [{\bf x}^i {\bf x}^j]_{\rm TF}  &=& \sum_{A, B}= \frac{Gm_B}{r^3} \left[\left\{2({\bf v}_A\times {\bf S}_A)\cdot {\bf r} -4({\bf v}_B\times {\bf S}_A)\cdot {\bf r}\right\} {\bf x}^i_A{\bf x}^j_A  \nonumber \right. \\ & & \left. - 4r^2 \left(({\bf v}_A\times {\bf S}_A)^i {\bf x}_A^j +({\bf v}_B\times {\bf S}_A)^i {\bf x}_A^j\right) \right]_{\rm STF},
\eea
and
\beq
\int  d^3 {\bf x} \dot T^{0k}_{\ref{nonlS}b} {\bf x}^k [{\bf x}^i{\bf x}^j]_{\rm TF} = \frac{d}{dt} \sum_{A, B} \frac{G m_B}{r^3} \left[-2r^2({\bf x}_A\times{\bf S}_A)^i{\bf x}_A^j - \frac{1}{2} ({\bf x}_B\times{\bf S}_A)\cdot {\bf x}_A {\bf x}_A^i {\bf x}_A^j\right]_{\rm STF}.
\eeq

Finally Fig. \ref{nonlS}$c$ yields
\bea
& & \int d^3 {\bf x}\left(T^{00}+T^{ii}\right)_{\ref{nonlS}c} [{\bf x}^i {\bf x}^j]_{\rm TF} = \\
& & \sum_{A, B} \frac{4Gm_B}{r}\left[({\bf v}_B\times {\bf S}_A)^i({\bf x}^j_A+{\bf x}^j_B)+(1-\kappa)({\bf v}_A\times {\bf S}_A)^i{\bf x}^j_A+({\bf r}\times{\bf S}_A)^i\left({\bf v}_B^j-{\bf v}_A^j-\frac{{\bf v}_B\cdot {\bf r}}{r^2}({\bf x}_A^j+{\bf x}_B^j)\right) \right. \nn \\ & & \left. +
\frac{1}{2r^2}({\bf v}_A\times {\bf S}_A)\cdot {\bf r}~\left\{(1+\kappa){\bf x}_B^i{\bf x}_B^j +(\kappa-1){\bf x}_A^i{\bf x}_A^j\right\}\right]_{\rm STF}, \nn
\eea
and
\bea
\int  d^3 {\bf x}~\dot T^{0k}_{\ref{nonlS}c}{\bf x}^k [{\bf x}^i{\bf x}^j]_{\rm TF}  &=&  \frac{d}{dt} \sum_{A, B} \frac{Gm_B}{r^3}\left[  r^2\left( \frac{1}{2}({\bf r}\times {\bf S}_A)^i {\bf r}^j+({\bf x}_B\times {\bf S}_A)^i ({\bf x}_A^j+{\bf x}_B^j)\right)\right. \\ & & \left. - ({\bf x}_A\times{\bf S}_A)\cdot {\bf x}_B {\bf x}_B^i{\bf x}_B^j-{\bf r}\cdot{\bf x}_B~({\bf r}\times{\bf S}_A)^i({\bf x}^j_A+{\bf x}_B^j) \right]_{\rm STF}. \nn
\eea

Combining these ingredients we have in the covariant SSC
\bea
& & I^{ij}_{\ref{nonlS}abc} = \sum_{A, B} \frac{2Gm_B}{r^3} \left[ ({\bf v}_B\times {\bf S}_A)\cdot {\bf r}({\bf x}^i_B{\bf x}^j_B-2{\bf x}_A^i{\bf x}_A^j)+({\bf v}_A\times {\bf S}_A)\cdot {\bf r}({\bf x}^i_A{\bf x}^j_A+{\bf x}_B^i{\bf x}_B^j)\right. \nn  \\ & & \left. + 2r^2 \left\{({\bf v}_B\times {\bf S}_A)^i {\bf x}_B^j -({\bf v}_A\times {\bf S}_A)^i {\bf x}_A^j+({\bf r}\times{\bf S}_A)^i\left({\bf v}_B^j-{\bf v}_A^j-\frac{{\bf v}_B\cdot {\bf r}}{r^2}({\bf x}_A^j+{\bf x}_B^j)\right)\right\} \right]_{\rm STF} \nn \\ & &  - \frac{4}{3}  \sum_{A, B} \frac{d}{dt} \left[ \frac{Gm_B}{r^3} \left\{r^2\left( \frac{1}{2} ({\bf x}_B\times {\bf S}_A)^i{\bf x}_A^j-\frac{3}{2}({\bf x}_A\times {\bf S}_A)^i{\bf x}_A^j +\frac{3}{2} ({\bf x}_B\times {\bf S}_A)^i{\bf x}_B^j-\frac{1}{2}({\bf x}_A\times {\bf S}_A)^i{\bf x}_B^j \right) \right.\right. \nn \\ & &\left. \left. - {\bf r}\cdot{\bf x}_B~({\bf r}\times{\bf S}_A)^i({\bf x}^j_A+{\bf x}_B^j)+ \frac{1}{2} ({\bf x}_A\times{\bf S}_A)\cdot {\bf x}_B ({\bf x}_A^i{\bf x}_A^j-2{\bf x}_B^i{\bf x}_B^j)\right\}\right]_{\rm STF}
\label{Qij5abc}.
\eea

\subsection{Current Quadrupole}

\subsubsection{Worldline radiation}

The current quadrupole $J^{ij}$ at $\mathcal O(\mathbf S_A)$ is needed to NLO. The components of $T^{\mu \nu}$ at $\mathcal O(\mathbf S_A)$ from worldline radiation including all subleading terms were already given above in Eqs. (\ref{set}) or (\ref{eq:WLrad00SA} - \ref{eq:WLradijSA}). Expanding these to obtain the moments of $T^{\mu \nu}$ in the current quadrupole $J^{ij}$ of Eqs. (\ref{J0ij}, \ref{J1ij}), we find for the worldline contributions to the current quadrupole from the diagram in Fig.~\ref{world}$a$ 
\bea
J^{ij}_{\ref{world}a} &=& \sum_A \left[\left(\frac{\kappa}{2}-\frac{3}{14}\right){\bf S}_A\cdot {\bf x}_A {\bf a}_A^i{\bf x}_A^j+\frac{1}{7}{\bf x}_A^i{\bf x}_A^j {\bf a}_A\cdot {\bf S}_A+\left(\frac{1}{14}-\frac{\kappa}{2}\right){\bf S}_A\cdot {\bf v}_A {\bf v}^i_A{\bf x}_A^j\right. \nonumber \\ & & \left. 
+\frac{11}{28} {\bf S}_A^i{\bf a}_A^j {\bf x}_A^2+{\bf S}_A^i {\bf x}_A^j\left(\frac{3}{2}-\left(\frac{3}{14}-\frac{\kappa}{2}\right) {\bf v}_A^2-\left(\frac{\kappa}{2}+\frac{3}{14}\right) {\bf a}_A\cdot {\bf x}_A\right)\right. \nonumber \\ & & \left. +
\left(\frac{4}{7}-\kappa\right){\bf v}^i_A{\bf S}_A^j {\bf v}_A\cdot {\bf x}_A +\left(\kappa-\frac{3}{14}\right){\bf S}_A\cdot {\bf x}_A {\bf v}_A^i{\bf v}_A^j\right]_{\rm STF}, 
\eea
which takes the following form in the covariant SSC 
\bea
\label{Jijsglin}
J^{ij}_{\ref{world}a} &=& \sum_A \left[ {\bf S}_A^i {\bf x}_A^j\left(\frac{3}{2} + \frac{2}{7} {\bf v}_A^2-\frac{5}{7} {\bf a}_A\cdot {\bf x}_A\right)-
\frac{3}{7} {\bf v}^i_A{\bf S}_A^j {\bf v}_A\cdot {\bf x}_A +\frac{11}{28} {\bf S}_A^i{\bf a}_A^j {\bf x}_A^2\right. \nonumber \\ & & \left. 
+\frac{2}{7} {\bf S}_A\cdot {\bf x}_A {\bf a}_A^i{\bf x}_A^j+\frac{1}{7} {\bf x}_A^i{\bf x}_A^j {\bf a}_A\cdot {\bf S}_A-\frac{3}{7}{\bf S}_A\cdot {\bf v}_A {\bf v}^i_A{\bf x}_A^j+\frac{11}{14} {\bf S}_A\cdot {\bf x}_A {\bf v}_A^i{\bf v}_A^j\right]_{\rm STF} .
\eea

The worldline radiation contribution from the diagram in Fig.~\ref{worldc}a to the current quadrupole $J^{ij}$ at $\mathcal O(\mathbf S_A^2)$ is just needed at LO and follows from Eqs. (\ref{eq:T0iSAsq}). This yields 
\beq
\label{Jijfslin}
 J_{\ref{worldc}a}^{ij} = \sum_A \frac{C_{ES^2}^{(A)}}{m_A} \left[ ({\bf v}_A\times {\bf S}_A)^i {\bf S}_A^j\right]_{\rm STF}.
\eeq

\subsubsection{Nonlinear gravitational effects}

For the nonlinear gravitational contributions to the current quadrupole we only need to compute the diagrams in Figs. \ref{nonlS}$bc$, since the contribution from \ref{nonlS}$a$ can be shown to be subleading. In order to calculate the $T^{0i}$ that enters in Eq. (\ref{J0ij}) we are only interested in diagrams with an external $\bar h_{0i}$. Notice that we already performed this computation for the NLO contributions linear in spin for the mass quadrupole moment. Thus the results for the current quadrupole from these diagrams are
\beq
 J^{ij}_{\ref{nonlS}b}= \sum_{A, B}\frac{G m_B}{2r^3}\left[{\bf S}_A \cdot {\bf x}_A {\bf r}^i{\bf x}_A^j-{\bf x}^i_A{\bf S}^j_A(6r^2+{\bf x}_A\cdot {\bf r}) \right]_{\rm STF}
\eeq
and
\bea
J^{ij}_{\ref{nonlS}c}&=& \sum_{A, B} \frac{Gm_B}{2r^3}\left[-3 {\bf S}_A\cdot {\bf x}_B ({\bf x}_A^i{\bf x}_A^j-{\bf x}_B^i{\bf x}_B^j)+{\bf S}_A\cdot {\bf x}_A ({\bf x}_A^i{\bf x}_A^j+2{\bf x}_A^i{\bf x}_B^j-3{\bf x}_B^i{\bf x}_B^j)\right. \nn \\ & & \left. +2{\bf S}_A^i{\bf x}_A^j {\bf x}_A\cdot {\bf r} \right]_{\rm STF},
\eea
which  add up to  
\bea
J^{ij}_{\ref{nonlS}bc}&=& \sum_{A, B} \frac{Gm_B}{2r^3}\left[
{\bf S}_A\cdot {\bf x}_A \left(2{\bf x}_A^i{\bf x}_A^j+{\bf x}_A^i{\bf x}_B^j-3{\bf x}_B^i{\bf x}_B^j\right)\nonumber \right. \\ & & \left.+3 {\bf S}_A\cdot {\bf x}_B ({\bf x}_B^i{\bf x}_B^j-{\bf x}_A^i{\bf x}_A^j) +{\bf S}_A^i{\bf x}_A^j ({\bf x}_A\cdot {\bf r}-6r^2) \right]_{\rm STF}.
\label{Jij5ab}
\eea

\subsection{Mass octupole}

For the mass octupole we only need the pieces linear in $\bar h_{\mu\nu}$ coming from the worldline couplings. Using  Eqs. (\ref{Q0ijk}, \ref{Q1ijk}) and (\ref{set}) we obtain
\beq
I_{\ref{world}a}^{ijk} = \sum_A \left[ \left(\frac{3}{2}+3\kappa\right)({\bf v}_A\times {\bf S}_A)^i{\bf x}_A^j {\bf x}_A^k -3 ({\bf x}_A\times {\bf S}_A)^i{\bf v}_A^j {\bf x}_A^k \right]_{\rm STF} \label{Q1ijklin}
\eeq
and
\beq
\label{Qijkfslin}
I^{ijk}_{\ref{worldc}a} = -\sum_A \frac{3C_{ES^2}^{(A)}}{m_A} [{\bf S}^i_A{\bf S}_A^j {\bf x}_A^k]_{\rm STF}.
\eeq

\subsection{Current octupole}

Similarly to the mass octupole, here we only need the linearized worldline coupling. Using  Eqs. (\ref{Jijk}) and (\ref{set}) we obtain
\beq
\label{Jijksglin}
J^{ijk}_{\ref{world}a}=2 \sum_A \left[ {\bf x}_A^i {\bf x}_A^j {\bf S}_A^k\right]_{\rm STF}.
\eeq

\section{Towards spin effects in the energy flux to 3PN} \label{sec:total}

Let us  start by collecting the different contributions to the multipole moments in Eqs. (\ref{Q012ijnlo}, \ref{Qijfslin}, \ref{3ab}, \ref{Qij4abc}, \ref{Qij5abc}, \ref{Jijsglin}, \ref{Jijfslin}, \ref{Jij5ab}, \ref{Qijkfslin}, \ref{Jijksglin}). In the covariant SSC the results are 
\bea
\label{Qijtot}
& & I^{ij}_{{\bf S}_A,{\bf S}_A^2,{\bf S}_A{\bf S}_B}= \sum_A \left[\frac{8}{3} ({\bf v}_A\times {\bf S}_A)^i {\bf x}_A^j  -\frac{4}{3} ({\bf x}_A\times {\bf S}_A)^i{\bf v}_A^j-\frac{4}{3}({\bf x}_A\times \dot {\bf S}_A)^i{\bf x}_A^j\right.  \\ 
& & -\left.  \frac{4}{3}\frac{d}{dt} \left\{{\bf v}_A \cdot {\bf x}_A({\bf v}_A\times{\bf S}_A)^i{\bf x}_A^j\right\}+\frac{1}{7} \frac{d^2}{dt^2}\left\{\frac{1}{3} {\bf x}_A\cdot{\bf v}_A  ({\bf x}_A\times{\bf S}_A)^i{\bf x}_A^j\right. \right. \nn \\ & & +
\left. \left. 4 {\bf x}_A^2 ({\bf v}_A\times{\bf S}_A)^i{\bf x}_A^j+{\bf x}_A^2 ({\bf S}_A\times{\bf x}_A)^i{\bf v}_A^j-\frac{5}{6}({\bf v}_A\times{\bf S}_A)\cdot {\bf x}_A~{\bf x}_A^i{\bf x}_A^j \right\} \right]_{\rm STF} \nn \\
& & + \sum_{A, B} \frac{2Gm_B}{r^3} \left[ ({\bf v}_B\times {\bf S}_A)\cdot {\bf r}({\bf x}^i_B{\bf x}^j_B-2{\bf x}_A^i{\bf x}_A^j)+({\bf v}_A\times {\bf S}_A)\cdot {\bf r}({\bf x}^i_A{\bf x}^j_A+{\bf x}_B^i{\bf x}_B^j)\right. \nn  \\ & & \left. + 2r^2\left\{ ({\bf v}_B\times {\bf S}_A)^i ({\bf x}_B^j -{\bf x}_A^j) +({\bf r}\times{\bf S}_A)^i\left({\bf v}_B^j-{\bf v}_A^j-\frac{{\bf v}_B\cdot {\bf r}}{r^2}({\bf x}_A^j+{\bf x}_B^j)\right)\right\} \right]_{\rm STF} \nn \\ & &  - \frac{2}{3} \sum_{A, B} \frac{d}{dt} \left[ \frac{Gm_B}{r^3} \left\{r^2\left( ({\bf x}_B\times {\bf S}_A)^i{\bf x}_A^j-3({\bf x}_A\times {\bf S}_A)^i{\bf x}_A^j +3 ({\bf x}_B\times {\bf S}_A)^i{\bf x}_B^j-({\bf x}_A\times {\bf S}_A)^i{\bf x}_B^j \right) \right.\right. \nn \\ & &\left. \left. - 2{\bf r}\cdot{\bf x}_B~({\bf r}\times{\bf S}_A)^i({\bf x}^j_A+{\bf x}_B^j)+ ({\bf x}_A\times{\bf S}_A)\cdot {\bf x}_B ({\bf x}_A^i{\bf x}_A^j-2{\bf x}_B^i{\bf x}_B^j)\right\}\right]_{\rm STF} \nn \\
& & + \sum_A \frac{C_{ES^2}^{(A)}}{m_A}\left[{\bf S}_A^i{\bf S}_A^j\left(-1+\frac{13}{42} {\bf v}_A^2+\frac{17}{21} {\bf a}_A\cdot {\bf x}_A\right)+{\bf S}_A^2\left(-\frac{11}{21}{\bf v}^i_A{\bf v}^j_A +\frac{10}{21}  {\bf a}_A^i{\bf x}_A^j\right)\right. \nn \\ & & - \left.  \frac{8}{21}{\bf x}^i_A{\bf S}_A^j {\bf a}_A\cdot {\bf S}_A+\frac{4}{7}{\bf v}^i_A{\bf S}_A^j{\bf S}_A\cdot {\bf v}_A-\frac{22}{21}{\bf a}_A^i{\bf S}_A^j {\bf S}_A\cdot {\bf x}_A\right]_{\rm STF} \nn \\ & & + \sum_{A, B} \frac{G}{2r^3}\left[\frac{C_{ES^2}^{(B)}m_A}{m_B}\left({\bf S}_B^2+9({\bf S}_B\cdot {\bf n})^2\right){\bf x}_B^i{\bf x}_B^j+6\frac{C_{ES^2}^{(B)}m_A}{m_B} r^2 {\bf S}_B^i{\bf S}_B^j \nonumber \right. \\ & & + \left.\left(\frac{C_{ES^2}^{(B)}m_A}{m_B}\left(3({\bf S}_B\cdot {\bf n})^2-{\bf S}_B^2\right)+12{\bf S}_A\cdot {\bf n} {\bf S}_B\cdot {\bf n}-4{\bf S}_A\cdot {\bf S}_B \right){\bf x}_A^i{\bf x}_A^j\right. \nonumber \\ & & - \left.  4 \frac{C_{ES^2}^{(B)}m_A}{m_B} {\bf S}_B^2{\bf x}_A^i{\bf x}_B^j+4\left( 3\frac{C_{ES^2}^{(B)}m_A}{m_B}{\bf S}_B\cdot {\bf r}+2 {\bf S}_A\cdot {\bf r}\right){\bf S}_B^i{\bf x}_B^j\right]_{\rm STF}\nn 
\eea
\bea
\label{Jijtot}
J^{ij}_{{\bf S}_A,{\bf S}_A^2}&=&  \sum_A \left[ \frac{C_{ES^2}^{(A)}}{m_A}  ({\bf v}_A\times {\bf S}_A)^i {\bf S}_A^j+
{\bf S}_A^i {\bf x}_A^j\left(\frac{3}{2} + \frac{2}{7} {\bf v}_A^2-\frac{5}{7} {\bf a}_A\cdot {\bf x}_A\right)-
\frac{3}{7} {\bf v}^i_A{\bf S}_A^j {\bf v}_A\cdot {\bf x}_A \right.  \\ & & \left. 
+\frac{11}{28} {\bf S}_A^i{\bf a}_A^j {\bf x}_A^2+\frac{2}{7} {\bf S}_A\cdot {\bf x}_A {\bf a}_A^i{\bf x}_A^j+\frac{1}{7} {\bf x}_A^i{\bf x}_A^j {\bf a}_A\cdot {\bf S}_A-\frac{3}{7}{\bf S}_A\cdot {\bf v}_A {\bf v}^i_A{\bf x}_A^j+\frac{11}{14} {\bf S}_A\cdot {\bf x}_A {\bf v}_A^i{\bf v}_A^j\right]_{\rm STF} \nonumber \\
 &&+\sum_{A, B}\frac{Gm_B}{2r^3}\left[3 {\bf S}_A\cdot {\bf x}_B ({\bf x}_B^i{\bf x}_B^j-{\bf x}_A^i{\bf x}_A^j)+ {\bf S}_A\cdot {\bf x}_A (2{\bf x}_A^i{\bf x}_A^j+{\bf x}_A^i{\bf x}_B^j-3{\bf x}_B^i{\bf x}_B^j)\nonumber \right. \\ & & \left. +{\bf S}_A^i{\bf x}_A^j ({\bf x}_A\cdot {\bf r}-6r^2) \right]_{\rm STF} \nn \\ \label{Joctup}
J^{ijk}_{{\bf S}_A} &=& 2 \sum_A \left[ {\bf x}_A^i {\bf x}_A^j {\bf S}_A^k\right]_{\rm STF}\\
\label{Qoctup}
I^{ijk}_{{\bf S}_A, {\bf S}_A^2} &=& \sum_A \left[ \frac{9}{2}({\bf v}_A\times {\bf S}_A)^i{\bf x}_A^j {\bf x}_A^k -3({\bf x}_A\times {\bf S}_A)^i{\bf v}_A^j {\bf x}_A^k -3\frac{C_{ES^2}^{(A)}}{m_A} {\bf S}^i_A{\bf S}_A^j {\bf x}_A^k \right]_{\rm STF},
\eea
plus the spin-independent contributions (see Appendix \ref{app:C})
\bea
\label{Qijsless}
I^{ij}_{\not {\bf S}} &=& \sum_A  m_A \left[ \left( 1+ \frac{3}{2} {\bf v}_A^2- \sum_{B} \frac{G m_B}{r}\right){\bf x}_A^i{\bf x}_A^j + \frac{11}{42} \frac{d^2}{dt^2}\left\{{\bf x}_A^2 {\bf x}_A^i{\bf x}_A^j \right\}  
-\frac{4}{3} \frac{d}{dt}\left\{ {\bf x}_A\cdot{\bf v}_A {\bf x}_A^i{\bf x}_A^j\right\}\right]_{\rm TF}  \\
J_{\not{\bf S}}^{ij} &=&  \sum_A m_A \left(1+\frac{{\bf v}_A^2}{2}\right) \left[({\bf x}_A\times{\bf v}_A)^i{\bf x}_A^j\right]_{\rm STF} + \sum_{A, B} \frac{Gm_Am_B}{r} \left[ 2({\bf x}_A\times{\bf v}_A)^i{\bf x}_A^j  \right.  \\ & &\left.  -\frac{11}{4}({\bf x}_B\times{\bf v}_A)^i{\bf x}_B^j-\frac{3}{4}({\bf x}_B\times{\bf v}_A)^i{\bf x}_A^j +({\bf x}_A\times{\bf v}_A)^i{\bf x}_B^j+\frac{7}{4}({\bf x}_A\times{\bf x}_B)^i{\bf v}_A^j\right.  \nn \\ & & \left.  +
\frac{{\bf v}_A\cdot{\bf r}}{4r^2} ({\bf x}_A\times{\bf x}_B)^i({\bf x}_A^j+{\bf x}_B^j)\right]_{\rm STF} + \frac{1}{28}\frac{d}{dt}\left[ \sum_A  m_A({\bf x}_A \times {\bf v}_A)^i (3 {\bf x}_A^2 {\bf v}_A^j-{\bf x}_A\cdot {\bf v}_A {\bf x}_A^j) \right. \nn \\ & & + \left.  \sum_{A, B} \frac{Gm_Am_B}{2r^3}  {\bf x}_A^i ({\bf x}_A\times{\bf x}_B)^j(6{\bf x}_A^2 -7 {\bf x}_A \cdot {\bf x}_B +7{\bf x}_B^2 )\right]_{\rm STF} \nn \\
J^{ijk}_{\not{\bf S}} &=&  \sum_A m_A \left[({\bf x}_A\times{\bf v}_A)^i{\bf x}_A^j{\bf x}^k_A\right]_{\rm STF}\\
I^{ijk}_{\not{\bf S}} &=& \sum_A m_A \left[ {\bf x}_A^i{\bf x}_A^j{\bf x}_A^k\right]_{\rm TF}.
\label{octsless}
\eea

The spin dependence of the multipoles is given in the covariant SSC which was used in \cite{nrgrss,nrgrs2,nrgrso} to compute the NLO spin(1)spin(2), spin(1)spin(1) and spin-orbit potentials. 
We could combine the results also in the NW SSC, $\kappa=1/2$. This would simplify the calculation of the power loss for spin(1)spin(2) and spin-orbit effects by using the results in \cite{eih,comment,nrgrss,nrgrso}. However, the NW SSC ceases to be useful once we include spin(1)spin(1) (finite size) corrections, where we found the covariant SSC to be more convenient. In any case, we can always transform the spin dynamics into precession form at the end of the day by means of a spin (and coordinate) redefinition \cite{nrgr, nrgrss,nrgrs2,nrgrso}. Ultimately the power loss will not depend on the choice of coordinates but its specific form will depend on the choice of spin variables. 

Our results in Eqs. (\ref{Qijtot} - \ref{octsless}) are the last missing ingredients for the spin-dependent parts of the energy flux to 3PN. Its computation will be presented in \cite{powerloss}. Besides the multipole moments in Eqs. (\ref{Qijtot}-\ref{octsless}) we also need to make use of a) the conservative spin dynamics to NLO  \cite{nrgrs,eih,comment,nrgrss,nrgrs2} where appropriate, and b) the modification of the energy flux linear in spin due to the leading tail effect \cite{rad1}. Furthermore, we can also use the multipole moments presented here in the computation of the gravitational waveform $h_{ij}^{\rm TT}$.

\section{Conclusions}\label{sec:concl}

In this paper we computed the necessary multipole moments, expressed in Eqs. (\ref{Qijtot} - \ref{octsless}) in the covariant SSC, which together with the results reported in \cite{nrgr, nrgrs,nrgrss,nrgrs2,rad1,nrgrso} will lead to the spin-orbit and spin-spin contributions to the power loss to 3PN order via standard procedures. For the computation we employed the EFT techniques developed in \cite{nrgr,nrgrs,rad1}. There is clearly no stumbling block to continue this program to higher orders in the PN expansion. We will report the final expressions for the energy flux and the phase in a separate paper \cite{powerloss}.

\begin{center}
{\bf Acknowledgments}
\end{center}

This work was supported in part by NSF under grants No.04-56556 and AST-0807444, and NASA grant NNX10AH14G (RAP); the US Department of Energy contract DE-FG-02-92ER40704 (AR), and NASA grant 22645.1.1110173 (IZR). We thank the Aspen Center for Physics (AR) and the Cal-Tech high energy theory group (RAP \& IZR) for their hospitality during the final stages of this work. IZR also acknowledges support from the Gordon and Betty Moore foundation.

\appendix

\section{Contribution from the term responsible for the preservation of the SSC}\label{app:B}

To describe spin dynamics in GR, and ensure the preservation of the SSC upon evolution, in \cite{nrgrproc,nrgrss,nrgrs2} the following Routhian was introduced (up to finite size effects) 
\beq
\label{actR}
{\cal R} =-\sum_A \left( m_A \sqrt{u^2_A} + 
\frac{1}{2}S_{A}^{ab}\omega_{ab\mu} u^\mu_A +\frac{1}{2m_A}R_{d
e a b}(x_A){S_{Ac}}^d S^{a b}_{A} \frac{u^e_A u_A^c}{\sqrt{u^2}}
+\cdots\right),
\eeq
where the ellipses represent nonlinear terms in the curvature necessary to account for the mismatch between $p$ and $u$ once the SSC is enforced. Since at 3PN order we can consider the covariant SSC to be $S^{ab}u_b=0$, we can ignore the higher order terms. The equations of motion follow from
\beq
\frac{\delta }{\delta x^\mu}\int {\cal R} d\lambda=0, \;\;\; \frac{d
S^{ab}}{d\lambda} = \{S^{ab},{\cal R}\}\label {eomV},
\eeq
and the Riemann dependent term in the Routhian guarantees the SSC is preserved upon evolution \cite{nrgrss,nrgrs2}. Notice that it vanishes once the SSC is enforced, however, it gives a nonzero contribution to the equations of motion from Eqs. (\ref{eomV}). In other words, we cannot set the SSC to zero until after we perform the variation of the Routhian. In particular,  
\beq
p^{\alpha}= \frac{1}{\sqrt{u^2}}\left(m
u^{\alpha}+\frac{1}{2m}R_{\beta\nu\rho\sigma}S^{\alpha\beta}S^{\rho\sigma}u^{\nu}\right)\label{up}.
\eeq

Naively it appears as if this extra term in the Routhian does not contribute to the radiation sector, since the SSC is conserved in time. However, we know it has to contribute to the stress-energy tensor, from which the equations of motion derive. Once again the subtle point lies in the fact that we cannot apply the SSC before we expand the metric field in terms of the radiation and potential modes. We notice that only the term where the radiation field stems from the vierbein inside the SSC contributes to the one-point function (see Eq. \ref{Srad}), and it reads
\beq
 -\sum_A \frac{1}{2m_A}R_{d
e a b}(H_{\mu\nu}(x_A))S^{a b}_{A} \frac{u^e_A}{\sqrt{u^2}} S^{cd}_{A} \frac{1}{2} {\bar h}_{\mu c} u_A^\mu,
\label{drsu}
\eeq
where we used 
\beq
e_\mu^c =  \delta^c_\mu + \frac{1}{2} {\bar h}_\mu^c + \cdots
\eeq  
All terms besides the one in Eq. (\ref{drsu}) which result from expanding the third term in Eq. (\ref{actR}) to linear order in the radiation field are proportional to the SSC and do not contribute.

The expression in (\ref{drsu}) produces a $\partial\partial H{\bar h}$ type of coupling which leads to a diagram equivalent to Fig. \ref{nonlS2}$b$. By standard power counting it is easy to show that this extra piece contributes an ${\cal O}({\bf S}_A^2)$ term in the mass quadrupole moment at 4PN.

Notice that the coupling in (\ref{drsu}) leads to the following expression for the stress tensor (after adding the spinless part from the worldline)
\beq
T^{\mu\nu}_{\rm \not{\bf S}+RSSuu} = \sum_A m_A u_A^\mu u_A^\nu +\frac{1}{2m_A}\langle R_{\beta\sigma\rho\alpha}\rangle(x_A) u_A^{(\nu} S_A^{\mu)\beta}S_A^{\rho\alpha}u_A^{\sigma}\equiv \sum_A p_A^{(\mu}u_A^{\nu)}, \label{Tup}
\eeq
where the expectation value of the Riemann tensor, i.e. $\langle R_{\beta\sigma\rho\alpha}\rangle$, is obtained in the background field of the companion. This is in accordance with similar expressions found in the literature (see for instance \cite{buo1,buo2}.)

\section{Spin-independent multipole moments}\label{app:C}

In this Appendix we summarize (within our conventions) the spin-independent multipole moments necessary to compute the radiated power linear in spin to 3PN order, some of which have been already reported in \cite{rad1}.\\

\begin{figure}[h!]
   \centering
    \includegraphics[width=10cm]{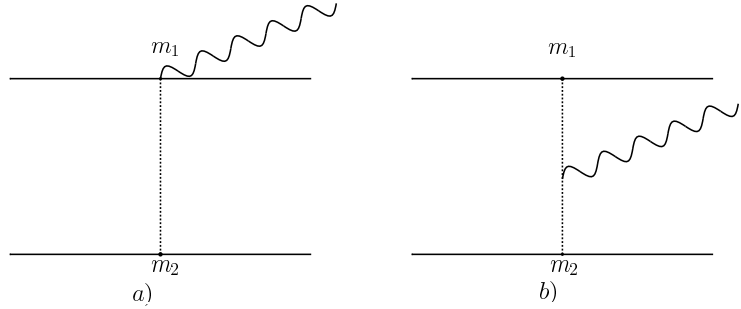}
\caption{\sl Diagrams representing the spin-independent nonlinear gravitational contributions to the multipole moments.} \label{segbenzJ}
\end{figure}

For the mass quadrupole (see Eq. (101) in \cite{rad1})
\bea
I^{ij}_{0+1}&=& \sum_A  m_A \left[ \left( 1+ \frac{3}{2} {\bf v}_A^2\right){\bf x}_A^i{\bf x}_A^j + \frac{11}{42} \frac{d^2}{dt^2}\left\{{\bf x}_A^2 {\bf x}_A^i{\bf x}_A^j \right\}  \right. \nn \\ & & \left.
-\frac{4}{3} \frac{d}{dt}\left\{ {\bf x}_A\cdot{\bf v}_A {\bf x}_A^i{\bf x}_A^j\right\} \right]_{\rm TF}-\sum_{A, B} \frac{G m_A m_B}{r} \left[{\bf x}_A^i{\bf x}_A^j\right]_{\rm TF}
\eea 
to 1PN order. For the current quadrupole the LO term reads (see Eq. (89) in \cite{rad1})
\beq
J_0^{ij} =  \sum_A m_A \left[({\bf x}_A\times{\bf v}_A)^i{\bf x}_A^j\right]_{\rm STF},
\eeq
whereas for the NLO contributions the new ingredients follow from: the radiation off the worldline couplings 
\beq
 \int d^3{\bf x} \, T^{0k}_{\ref{world}a}{\bf x}^j{\bf x}^l =\sum_A m_A \frac{{\bf v}_A^2}{2}  {\bf v}_A^k {\bf x}_A^j {\bf x}_A^l,
\eeq
and the diagrams in Figs. \ref{segbenzJ}$ab$ with an external $\bar h_{0k}$
\beq
\int d^3{\bf x} \, T^{0k}_{\ref{segbenzJ}a}{\bf x}^j{\bf x}^l =\sum_{A, B} \frac{Gm_A m_B}{r}  {\bf v}_A^k {\bf x}_A^j {\bf x}_A^l,
\eeq
\bea
 \int d^3{\bf x} \, T^{0k}_{\ref{segbenzJ}b}{\bf x}^j{\bf x}^l  & = & -\sum_{A, B} \frac{2Gm_Am_B}{r} \left[{\bf v}_A^k {\bf x}_B^j {\bf x}_B^l + {\bf r}^k{\bf v}_A^j ({\bf x}_A+{\bf x}_B)^l \right]_{\rm S} \\
&  + & \sum_{A, B} \frac{Gm_Am_B}{6r}\left[ {\bf v}_A^k({\bf x}_B^j{\bf x}_B^l+{\bf x}_A^j{\bf x}_A^l+{\bf x}_A^j{\bf x}_B^l)+r^k\left(2{\bf v}_A^j{\bf x}_A^l+{\bf v}_A^j{\bf x}_B^l \right.\right. \nn \\ &  - & \left. \left. \frac{{\bf v}_A\cdot{\bf r}}{r^2}({\bf x}_A^j{\bf x}_A^l+{\bf x}_B^j{\bf x}_B^l+{\bf x}_A^j{\bf x}_B^l)\right)\right]_{\rm S} \nn .
\eea

We also need the contribution to the current quadrupole from Eq. (\ref{J1ij}) with the LO $T^{ij}$. The LO $T^{ij}$ is comprised of a worldline contribution \cite{rad1}
\beq
T^{ij}(t, \mathbf k) = \sum_A m_A \mathbf v_A^i \mathbf v_A^j e^{-i \mathbf k \cdot \mathbf x_A},
\eeq
as well as of a term from the diagram in Fig. \ref{segbenzJ}$b$ with an external $\bar h_{ij}$. It yields the moments of $T^{ij}$ for Eq. (\ref{J1ij}) 
\beq
\frac{1}{28}\left[\epsilon^{ilk} \int d^3{\bf x} \, \dot T^{lm}_{\ref{segbenzJ}b} {\bf x}^j {\bf x}^k {\bf x}^m\right]_{\rm STF} = \frac{1}{28}\frac{d}{dt} \frac{Gm_Am_B}{2r^3} \sum_{A, B} \left[ ({\bf x}_A\times {\bf x}_B)^i {\bf x}_A^j \left( -{\bf x}_A\cdot {\bf x}_B+{\bf x}_B^2 \right)\right]_{\rm STF}
\eeq
and
\beq
\frac{3}{28}\left[\epsilon^{ilk} \int d^3{\bf x} \, \dot T^{jk}_{\ref{segbenzJ}b} {\bf x}^l {\bf x}^2\right]_{\rm STF} = \frac{1}{28} \frac{d}{dt} \frac{Gm_Am_B}{2r^3} \sum_{A, B} \left[ ({\bf x}_A\times {\bf x}_B)^i{\bf x}_A^j \left(6{\bf x}_A^2+6{\bf x}_B^2-6{\bf x}_A\cdot {\bf x}_B\right) \right]_{\rm STF},
\eeq
so that the total contribution from Eq. (\ref{J1ij}) reads
\bea
J_1^{ij} &=&  \frac{d}{dt} \left[\sum_A \frac{m_A}{28}  ({\bf x}_A \times {\bf v}_A)^i (3 {\bf x}_A^2 {\bf v}_A^j-{\bf x}_A\cdot {\bf v}_A {\bf x}_A^j) \right. \nn \\ &+& \left. \sum_{A, B} \frac{m_Am_B G}{56r^3}  {\bf x}_A^i ({\bf x}_A\times{\bf x}_B)^j(6{\bf x}_A^2-7 {\bf x}_A \cdot {\bf x}_B +7{\bf x}_B^2)\right]_{\rm STF}.
\eea

Adding the pieces together to 1PN order
\bea
\label{J01ijnoS}
& & J_{0+1}^{ij} =  \sum_A m_A \left(1+\frac{{\bf v}_A^2}{2}\right) \left[({\bf x}_A\times{\bf v}_A)^i{\bf x}_A^j\right]_{\rm STF} + \sum_{A, B} \frac{Gm_Am_B}{r} \left[ 2({\bf x}_A\times{\bf v}_A)^i{\bf x}_A^j  \right.  \\ & &\left.  -\frac{11}{4}({\bf x}_B\times{\bf v}_A)^i{\bf x}_B^j-\frac{3}{4}({\bf x}_B\times{\bf v}_A)^i{\bf x}_A^j +({\bf x}_A\times{\bf v}_A)^i{\bf x}_B^j+\frac{7}{4}({\bf x}_A\times{\bf x}_B)^i{\bf v}_A^j\right.  \nn \\ & & \left.  +
\frac{{\bf v}_A\cdot{\bf r}}{4r^2} ({\bf x}_A\times{\bf x}_B)^i({\bf x}_A^j+{\bf x}_B^j)\right]_{\rm STF} + \frac{1}{28}\frac{d}{dt}\left[ \sum_A  m_A({\bf x}_A \times {\bf v}_A)^i (3 {\bf x}_A^2 {\bf v}_A^j-{\bf x}_A\cdot {\bf v}_A {\bf x}_A^j) \right. \nn \\ & & + \left.  \sum_{A, B} \frac{Gm_Am_B}{2r^3}  {\bf x}_A^i ({\bf x}_A\times{\bf x}_B)^j(6{\bf x}_A^2 -7 {\bf x}_A \cdot {\bf x}_B +7{\bf x}_B^2 )\right]_{\rm STF}. \nn
\eea

Notice that the last term in Eq. (\ref{J01ijnoS}) vanishes in the CM frame and therefore does not radiate. Finally for the octupole moments at LO 
\beq
I^{ijk}_0 = \sum_A m_A \left[ {\bf x}_A^i{\bf x}_A^j{\bf x}_A^k\right]_{\rm TF}
\eeq
and
\beq
J^{ijk}_0 = \sum_A m_A \left[({\bf x}_A\times{\bf v}_A)^i{\bf x}_A^j{\bf x}^k_A\right]_{\rm STF},
\eeq

As a check, one can show that Eq. (\ref{J01ijnoS}) reproduces the result in \cite{Blanchet:2002av} for the  radiated power that follows from the current quadrupole piece in Eq. (\ref{powerloss}).

\end{document}